\documentclass[acmsmall]{acmart}
\usepackage{multirow}
\usepackage{graphics}
\usepackage{graphicx}%\usepackage{graphicxsp}
\usepackage{subcaption}
\usepackage{amsthm}
\usepackage{amsmath}
\usepackage{color,colortbl}
\usepackage[flushleft]{threeparttable}
\usepackage[dvipsnames]{xcolor}

\usepackage{booktabs}
\usepackage{tabularx}
\usepackage{array}
\usepackage{calc}
\title{The Price of Interoperability: Exploring Cross-Chain Bridges and Their Economic Consequences}

\author{Yiyue Cao}
\authornote{The first two authors contributed equally and are listed alphabetically.}
\affiliation{
\institution{
The Hong Kong University of Science and Technology (Guangzhou)
}
\country{China}
}
\email{ycao948@connect.hkust-gz.edu.cn}

\author{Mingzhe Zheng}
\authornotemark[1]
\affiliation{
\institution{
The Hong Kong University of Science and Technology (Guangzhou)
}
\country{China}
}
\email{mzheng842@connect.hkust-gz.edu.cn}

\author{Lin William Cong}
\affiliation{
\institution{Nanyang Technological University}
\country{Singapore}
}
\email{will.cong@ntu.edu.sg}

\author{Siguang Li}
\affiliation{
\institution{
The Hong Kong University of Science and Technology (Guangzhou)
}
\country{China}
}
\email{siguangli@hkust-gz.edu.cn}

\author{Xuechao Wang}
\authornote{Corresponding author.}
\affiliation{
\institution{
The Hong Kong University of Science and Technology (Guangzhou)
}
\country{China}
}
\email{xuechaowang@hkust-gz.edu.cn}

\begin{document}

\setcopyright{cc}
\setcctype{by}
\acmJournal{POMACS}
\acmYear{2026}
\acmVolume{10}
\acmNumber{2}
\acmArticle{52}
\acmMonth{6}
\acmDOI{10.1145/3805650}

% % % % % % % % % % % % % % % % % % % % % % % % % % % % % % % % % % % % % % % % % % % % % % 
\begin{abstract}

Modern blockchain ecosystems consist of many heterogeneous networks, creating an essential need for interoperability that allows assets to move seamlessly across chains. Cross-chain bridges serve as the primary infrastructure enabling this interoperability by executing verifiable state transitions that reallocate assets and liquidity across networks. Although prior work has focused largely on bridge design and security, the system-level and economic consequences of liquidity interoperability remain poorly understood.

This paper presents a large-scale empirical measurement study of cross-chain interoperability based on a comprehensive dataset covering 20 blockchains and 16 major bridge protocols from 2022 to 2025. We model the multi-chain ecosystem as a time-varying weighted hypergraph and introduce metrics to characterize blockchain interoperability via bridges. 
We define two metrics of interoperability. Structural interoperability captures the connectivity implied by deployed bridge infrastructure, reflecting bridge coverage and redundancy independent of user behavior. Active interoperability captures how intensively users rely on cross-chain routing in practice, measured as realized cross-chain transfer volumes normalized by chain size. This decomposition enables analysis of infrastructure capacity and utilization as distinct dimensions.
Our analysis reveals several key findings. First, the cross-chain network evolves from a sparse hub-and-spoke topology into a dense, multi-hub core dominated by EVM-compatible chains. 
Second, we show that cross-chain capital flows are return-chasing and cost-sensitive, whereas structural interoperability primarily serves as an enabling factor rather than a directional driver of flows.
Third, we identify a growth-return paradox: while higher interoperability expands ecosystem scale, it dilutes native asset value. 
Fourth, we uncover an efficiency-fragility trade-off: structural interoperability acts as a load balancer that reduces average gas fees, but it also synchronizes economic cycles. Bridges transform independent networks into a tightly coupled system.
Finally, we show that architectural heterogeneity across bridge designs leads to distinct economic and system-level outcomes, yielding direct implications for the design of scalable and resilient cross-chain infrastructure.

\end{abstract}

\begin{CCSXML}
<ccs2012>
   <concept>
       <concept_id>10003033.10003079.10011704</concept_id>
       <concept_desc>Networks~Network measurement</concept_desc>
       <concept_significance>500</concept_significance>
       </concept>
   <concept>
       <concept_id>10010405.10010455.10010460</concept_id>
       <concept_desc>Applied computing~Economics</concept_desc>
       <concept_significance>300</concept_significance>
       </concept>
 </ccs2012>
\end{CCSXML}

\ccsdesc[500]{Networks~Network measurement}
\ccsdesc[300]{Applied computing~Economics}

\keywords{cross-chain bridges, blockchain interoperability, network measurement, bridge activity, decentralized finance, multi-chain ecosystems}

\maketitle

% % % % % % % % % % % % % % % % % % % % % % % % % % % % % % % % % % % % % % % % % % % % 
\section{Introduction}

Blockchain ecosystems are evolving into a multi-chain economy in which users, liquidity, and applications move and interact across heterogeneous execution environments.
Cross-chain bridges serve as the primary on-chain infrastructure enabling this mobility by mediating verifiable state transitions that reallocate assets across chains.
Bridge-mediated transfers routinely reach hundreds of millions of USD per day~\cite{defillama_bridges},
and cumulative volumes across major bridge networks exceed tens of billions of USD~\cite{wormholescan_upgrade_blog}.
Interoperability is therefore no longer a peripheral feature but an increasingly central substrate that reshapes how congestion, liquidity, and risk are distributed across blockchain systems.

Despite this growing importance, the system-wide and economic consequences of cross-chain interoperability remain poorly understood.
A large body of work studies bridges from the perspective of protocol design and security, focusing on correctness and attack surfaces~\cite{zamyatin2021sok,belchior2021survey,notland2025sok}.
A smaller empirical literature measures cross-chain transfers and routing, but often treats observed flows as a proxy for connectivity and impact~\cite{hu2024piecing,augusto2025xchaindatagen,subramanian2024benchmarking}. These perspectives are insufficient for understanding system-wide effects, because they conflate two different aspects of interoperability: the connectivity implied by deployed infrastructure and the interaction intensity revealed by user behavior.
This makes it hard to distinguish whether interoperability expands effective system capacity or mainly reallocates activity across chains.

To address this gap, we propose a measurement framework that formally decouples infrastructure from utilization.
We model the ecosystem as a time-varying weighted hypergraph, where bridges act as hyperedges connecting multiple chains simultaneously.
From this representation, we derive \emph{structural interoperability}, a metric capturing the topological connectivity provisioned by infrastructure, and \emph{active interoperability}, a metric capturing the realized intensity of cross-chain transfers.
Structural interoperability reflects what the infrastructure \emph{can} connect, whereas active interoperability reflects what users \emph{actually} connect through realized transfers.

We construct a unified, UTC-aligned, and schema-normalized multi-source dataset covering 20 chains and 16 bridge protocols over 2022--2025.
It integrates on-chain traces, bridge APIs, total value locked, and price series.
This dataset supports daily panel analysis at the chain level and at the directed corridor level.
Our measurements first characterize a set of stylized facts about how interoperability evolves in practice.
The observed corridor network densifies rapidly from 2022 to 2025 and forms a multi-hub core dominated by large EVM-compatible stacks, while many peripheral chains remain connected through a small number of gateway corridors.
Bridge activity also exhibits a persistent count--notional asymmetry: episodes dominated by many small transfers need not coincide with episodes dominated by large value mobility, and the dominant bridges differ across these two views.
These stylized facts motivate separating deployed connectivity from realized usage when characterizing interoperability.

Our empirical analysis reveals systematic trade-offs associated with interoperability:
\begin{enumerate}
    \item \textbf{Growth--Return Paradox.}
    Structural interoperability is associated with ecosystem expansion.
    Controlling for chain and time fixed effects and standard covariates, a one-unit increase in ASI under our normalization is associated with higher TVL, higher user activity, and more contract deployments, with magnitudes on the order of a few percent.
    At the same time, higher ASI is associated with lower native-token returns at longer horizons, including approximately 0.4\% lower 30-day cumulative returns and 1.1\% lower 100-day cumulative returns per ASI unit.
    We propose a supply-side composition hypothesis that is consistent with this pattern: improved bridging lowers frictions for importing exogenous hard assets, which can substitute native tokens in collateral and settlement roles.
    We view this mechanism as a testable explanation rather than a directly identified channel.

    \item \textbf{Efficiency--Fragility Trade-off.}
    Structural interoperability is associated with lower execution-cost proxies.
    A one-unit increase in ASI is associated with about 0.7\% lower average gas per transaction and with lower aggregate gas usage and fees, consistent with activity being routed toward lower-cost execution environments when more paths are available.
    However, interoperability is also associated with fragility and stronger coupling.
    Using the Multichain collapse in a difference-in-differences analysis, we find that exposed chains experienced an additional decline of about 0.77 log points in TVL relative to non-exposed chains. 
    Moreover, higher structural interoperability between a pair of chains is associated with higher TVL co-movement, with about a 0.4\% increase in correlation per unit, suggesting stronger economic synchronization.
\end{enumerate}

Finally, we show that these relationships vary systematically with execution-stack compatibility and bridge architecture. Official and third-party bridges serve different economic functions, and settlement mechanisms differ in how well they scale with usage. These patterns suggest that operators should manage utilization rather than only expand coverage, differentiate canonical migration routes from external liquidity entry points, and match settlement mechanisms to expected corridor load. Our contagion results also carry implications for bridge security. Structural connectivity propagates economic shocks even without direct exploits, and the bridge designs most sensitive to congestion are also those most exposed to security incidents, suggesting that flow controls serve as both efficiency and security infrastructure.

In summary, this paper makes four contributions:
\begin{enumerate}
    \item \textbf{A unified measurement dataset.}
    We construct and publicly release a multi-source, UTC-aligned dataset covering 2022--2025 that supports reproducible measurement at the chain-day and bridge--corridor-day levels.

    \item \textbf{A capacity--utilization decomposition.}
    We model bridges as hyperedges in a time-varying weighted hypergraph and introduce structural and active interoperability metrics that separately capture deployed connectivity and realized transfer intensity.
    This separation avoids conflating flows with infrastructure and enables like-for-like comparisons across bridges and chains.

    \item \textbf{Ecosystem-level stylized facts and associations.}  
    Using the above framework, we document network densification, endpoint roles, and persistent count--notional asymmetry, and quantify how structural versus active interoperability relate differently to ecosystem scale, execution-cost proxies, and cross-chain co-movement.
    We organize these patterns as the Growth--Return Paradox and the Efficiency--Fragility Trade-off.

    \item \textbf{Heterogeneity grounded in architecture.}
    We show that these relationships vary systematically by execution-stack compatibility and by bridge design, including verification and settlement mechanisms.
    The results provide guidance on where interoperability acts more like load balancing and where it is more associated with congestion and tighter coupling.
\end{enumerate}

% % % % % % % % % % % % % % % % % % % % % % % % % % % % % % % % % % % % % % % % % % % % 
\section{Background: Cross-Chain Bridges and Interoperability}
Blockchains are independent replicated execution environments.
Each chain maintains its own state, executes transactions under a specific virtual machine and smart-contract runtime, and reaches consensus on an ordered log of state transitions.
Smart contracts encode application logic, including token issuance, swaps, lending, and governance, and define the rules by which assets can be created, moved, and composed within a chain.
Because state and contract execution are not shared across chains, economic primitives are instantiated per chain rather than globally shared, including stablecoins, AMM pools, and collateral positions in lending protocols.
This per-chain instantiation fragments liquidity, induces heterogeneous execution costs, and exposes users to chain-specific finality and security conditions.

Bridging these fragmented execution domains requires cross-chain mechanisms with heterogeneous functionality and trust assumptions~\cite{zamyatin2021sok,belchior2021survey}.
In this work, we focus on the subset that leaves comparable on-chain traces, namely asset and liquidity mobility via bridges.
Cross-chain asset movement can also occur through centralized exchanges, where users deposit on one chain and withdraw on another after internal netting on the venue ledger.
This channel is economically important but largely opaque at the corridor level, because most of the transfer occurs off-chain and only deposits and withdrawals are externally visible.
Bridge routing, in contrast, leaves standardized on-chain traces at both endpoints and makes key execution frictions observable, including latency and fees that vary with verification and settlement design.
Accordingly, our measurement and empirical analysis center on bridge-mediated transfers, and we report results at three granularities: bridges, chains, and directed corridors defined by source--destination chain pairs.

% \subsection{The Cross-Chain State Mediator: Verification and Asset Settlement}
\paragraph{Bridge abstraction.}
We view a cross-chain bridge as a state-transition mediator that converts a source-chain commitment into a destination-chain settlement under an explicit verification rule.
A transfer can be represented as $\tau=(c_s,a,x,u_s,c_d,u_d)$: sender $u_s$ commits amount $x$ of asset $a$ on source chain $c_s$, and an economically corresponding position is materialized for recipient $u_d$ on destination chain $c_d$.

Conceptually, a bridge transfer follows a three-step pipeline.
First, \emph{initiation} occurs on $c_s$, where assets are locked, burned, or deposited into a bridge contract or liquidity pool, thereby specifying the asset type, amount, and redemption conditions.
Second, the bridge performs \emph{verification} by establishing that the source-chain event has occurred and is final enough to honor.
Depending on the design, verification may rely on a designated set of relayers that sign an authenticated message, on the publication of a compact commitment that summarizes source-chain events, or on a cryptographic proof that can be checked on the destination chain.
Third, \emph{settlement} occurs on $c_d$, where the recipient obtains the bridged position through minting or releasing an asset, or through a payout from local liquidity reserves, following the bridge's accounting rules.
User-facing bridging may additionally include route selection and destination execution, such as choosing among multiple bridges or combining a transfer with a swap, but the three-stage pipeline above captures the common on-chain traces most relevant for measurement~\cite{belchior2021survey,zamyatin2021sok}.

\paragraph{Design dimensions.}
Bridge designs primarily differ along two axes: \emph{asset settlement semantics} and \emph{verification}.
On the settlement side, common models include \textbf{Lock-and-Mint}, where assets remain in custody on $c_s$ while wrapped tokens are minted on $c_d$ and redeemability enforces a one-to-one backing;
\textbf{Burn-and-Mint}, where tokens are destroyed on $c_s$ and re-issued on $c_d$ to preserve a global supply constraint across instances; and
\textbf{Liquidity Pools}, where settlement is funded by local reserves on $c_d$, introducing liquidity availability constraints and potential slippage.
On the verification side, mechanisms range from trusted committees or oracle networks, to optimistic designs that accept messages unless challenged within a dispute window, to light-client or proof-based verification that reduces trust assumptions at higher operational and computational cost.

In \emph{Layer-1 (L1)}--\emph{Layer-2 (L2)} settings, a Layer-2 system provides additional execution capacity while using Layer-1 as the final settlement and security anchor.
Layer-2 is an umbrella term, and \emph{rollups} are the dominant design in which the Layer-2 executes transactions and maintains state off the Layer-1, then anchors state updates by posting periodic commitments to a smart contract on Layer-1.
This Layer-1 contract serves as the canonical bridge interface: it escrows deposits, records state commitments, and enforces withdrawals once the rollup update is accepted under its security mechanism, such as a challenge period in optimistic rollups or validity-proof verification in zk rollups~\cite{eth_optimistic_rollups,eth_zk_rollups}.
In contrast, Layer-1--Layer-1 or Layer-2--Layer-2 bridges connect systems with independent consensus and security assumptions, so verification must establish correctness and liveness across heterogeneous trust domains rather than through a single settlement anchor.

These design choices shape observable latency, fees, congestion sensitivity, and failure modes, motivating our later stratification of measurements by bridge mechanism and connectivity regime.

% % % % % % % % % % % % % % % % % % % % % % % % % % % % % % % % % % % % % % % % % % % % 

\section{Measurement Framework: A Hypergraph Representation of Interoperability}
\label{sec:measurement}

To quantify the systemic impact of cross-chain infrastructure, we must move beyond simple bilateral graph representations. Unlike pairwise channels, a bridge protocol typically supports a set of chains jointly; treating each corridor as an independent edge double-counts shared deployment decisions. We propose a hypergraph-based framework to formally decouple the \emph{structural capacity} of the infrastructure from its \emph{active utilization} by workloads.

\subsection{Network Representation: The Dynamic Multi-Chain Ecosystem}

We model the multi-chain ecosystem as a time-varying weighted hypergraph $\mathcal{H}_t = (\mathcal{V}, \mathcal{E}_t)$, constructed from daily snapshots of bridge deployments. Each vertex $v_i \in \mathcal{V}$ refers to blockchains, and a hyperedge $e_{k,t} \subseteq \mathcal{V}$ represents a bridge protocol $k$ that provides connectivity to a specific subset of chains at time $t$.

To quantify the structural importance of each bridge, we assign a time-varying weight $w_{k,t}$ to each hyperedge. Following our computational implementation, the weight is determined by the bridge's coverage breadth relative to the global ecosystem,
$w_{k,t} = \alpha \cdot \left( \frac{|e_{k,t}|}{N_{total}} \right)^\theta$,
where $|e_{k,t}|$ is the number of chains connected by bridge $k$, normalized by $N_{total}$, the total number of chains in the sample. $\alpha, \theta$ are scaling parameters. We set them equal to 1 as the benchmark. This weighting logic reflects that a bridge connecting a wider cluster of chains contributes greater structural breadth. 

We then project the hypergraph into a weighted graph over chains by aggregating pairwise connection strengths. For any two chains $i$ and $j$, the total connection strength is $S_{ij,t}=\sum_{i,j \in e_{k,t}} w_{k,t} (e_{k,t})$, i.e., the sum of weights of all bridges that connect both chains on day $t$. This strength is converted into a distance metric used for shortest-path computation: $\delta_{ij,t}=\frac{1}{1+S_{ij,t}}$, so that stronger connectivity implies lower traversal cost. We then compute shortest-path distances $d_{ij,t}$ between all chain pairs using Dijkstra’s algorithm on this weighted graph. The metric scheme is designed to capture the following features: (i) when more bridges connect two chains, the shortest distance of the pair of chains decreases; (ii) when a bridge has wider coverage, it contributes more to the distance; and (iii) the metrics are comparable across blockchains in the cross-section and over time.

\subsection{Metric Construction: Structural Capacity and Active Utilization}
\paragraph{1. Structural Metrics (Capacity).}
These metrics quantify the infrastructure provisioned by system architects, independent of user behavior.

    \textbf{Pairwise Structural Interoperability (PSI):} 
    PSI measures the aggregate friction-adjusted bandwidth between two specific chains, $i$ and $j$. It is the friction-adjusted proximity induced by the weighted projection and shortest paths.
    \begin{equation}
        \text{PSI}_{ij,t} = \frac{1}{2}\left( \frac{1}{d_{ij,t}} + \frac{1}{d_{ji,t}}  \right).
    \end{equation}
    A high $\text{PSI}_{ij,t}$ implies high redundancy and low theoretical friction for state transitions between the pair. This metric is the foundational unit for constructing the global topology.

    \textbf{Aggregated Structural Interoperability (ASI):} 
    ASI measures a chain's global integration. It is defined as the sum of reciprocal shortest paths to all other reachable nodes:
    \begin{equation}
        \text{ASI}_{i,t} = \sum_{j \neq i} \frac{1}{d_{ij,t}}.
    \end{equation}
    ASI captures how easily a given chain can reach all other chains through the bridge network. Chains that are connected to many others via multiple high-coverage bridges will have higher ASI, even if they do not currently experience large volumes of transfers. Importantly, ASI is driven by bridge deployment and coverage, and is therefore largely determined by system design rather than user behavior. In addition to the aggregate ASI, we also compute ASI separately for different bridge categories, such as official vs. third-party bridges, and lock-and-mint vs. burn-and-mint vs. liquidity-pool designs, by restricting the hyperedge set to the corresponding subsets before recomputing distances and centrality. This allows us to study how different bridge architectures contribute to structural connectivity and downstream economic outcomes.

\paragraph{2. Active Interoperability (utilization)}
Aggregated Active Interoperability (AAI) measures the \emph{Utilization} intensity. It is defined as the velocity of cross-chain capital flow relative to the local economy size.

Let $F_{i,t}^{\text{in}}$ and $F_{i,t}^{\text{out}}$ be the total gross USD inflows and outflows for chain $i$ across all bridges on day $t$. The AAI is calculated as:
\begin{equation}
    \text{AAI}_{i,t} = \frac{F_{i,t}^{\text{in}} + F_{i,t}^{\text{out}}}{\text{TVL}_{i,t}}.
\end{equation}
AAI measures how intensively a chain participates in cross-chain capital movement relative to its economic size.
Normalizing by TVL ensures that AAI reflects functional dependence on interoperability rather than absolute market scale.
Analogous to ASI, we also compute category-specific AAI measures by aggregating flows separately for different bridge types.

\subsection{Interpretation: The Capacity-Utilization Decoupling}

The distinction between structural and active interoperability is the measurement strategy that supports our empirical identification designs.
\begin{itemize}
    \item \textbf{Structural interoperability (capacity):} Represents the ``Pipes.'' It is a structural variable determined by engineering deployments and bridge integrations. It captures the \emph{potential} for interaction. Changes in ASI reflect changes in the ecosystem's topology.
    \item \textbf{Active interoperability (utilization):} Represents the ``Water.'' It is a behavioral variable determined by user incentives (e.g., arbitrage, yield farming). It captures the \emph{intensity} of interaction. Changes in AAI reflect changes in market demand.
\end{itemize}
By formally decoupling these dimensions, we can isolate the systemic effects of \emph{being connected} from the effects of \emph{interacting}. This allows us to test whether ecosystem outcomes—such as growth or risk contagion—are driven by the mere presence of infrastructure or by the volume of capital flowing through it.

% % % % % % % % % % % % % % % % % % % % % % % % % % % % % % % % % % % % % % % % % % % % 
\section{Data and Stylized Facts}

\subsection{Data Collection}
We build a multi-source measurement dataset for cross-chain interoperability by integrating
(i) on-chain activity and bridge traces, (ii) bridge-provider transfer records, and (iii) external series capturing chain size and valuation.
On-chain activity and a subset of bridge-mediated transfers are obtained from Dune, with extraction logic described in Appendix~\ref{app:dune}~\cite{dune}.
To improve corridor coverage for key bridges, we additionally collect transaction-level transfer records for Wormhole and deBridge (DLN) from their official APIs~\cite{wormholescan_api,dln_orders_api,dln_stats_api}.
To contextualize liquidity depth and valuation, we obtain chain- and protocol-level TVL time series from DefiLlama~\cite{defillama_chain_tvl_api,defillama_protocol_api},
and use native-token USD prices from Yahoo Finance when USD valuation requires token pricing, such as gas cost normalization~\cite{yfinance}.

All sources are aligned to a consistent UTC day boundary and normalized to a unified set of chain identifiers.
The resulting dataset spans 2022-01-01 to 2025-10-31 and provides two analysis-ready daily panels covering 20 chains and 16 bridge entities, summarized in Appendix Tables~\ref{tab:chains_covered} and~\ref{tab:bridges_covered}.
We additionally include auxiliary TVL and price series, and our coverage complements prior public measurement datasets that track fewer entities, such as 13 bridges across 7 blockchains~\cite{hu2024piecing}.
For bridges with API feeds, we collect 3.6M Wormhole transfer records and 5.4M deBridge DLN transfer records.
We reconcile these records to canonical on-chain logs by linking source and destination transaction hashes to Dune-indexed events, resolving token metadata such as contract address, decimals, and normalized amounts, and recomputing corridor-day aggregates under the same UTC boundary.
When upstream sources omit fields such as latency or fee components, we retain available measures and backfill only those quantities that can be reliably reconstructed.
We publicly release the cleaned dataset, code, and reproduction instructions in an online repository.\footnote{\url{https://github.com/Cyy0808/The-Price-of-Interoperability}}

\paragraph{Chain-day attributes.}
The first core panel is a chain-level daily attributes table, where each row corresponds to a chain $c$ on day $t$.
It summarizes adoption and activity, execution costs, and market context.
Variables include TVL measured in USD, developer and activity proxies such as \texttt{new\_contracts\_count} and \texttt{daily\_active\_users},
gas usage and expenditure such as \texttt{total\_gas\_used}, \texttt{total\_gas\_usd}, and \texttt{median\_gas\_usd},
and OHLCV-derived price and volume variables such as \texttt{Close\_Price\_USD} and \texttt{Volume\_USD}.
Complete definitions are provided in the data dictionary in the Appendix.

\paragraph{Bridge--corridor flows.}
The second core panel is a bridge--corridor daily flows table indexed by bridge $b$, source chain $i$, destination chain $j$, and day $t$.
Each row records transfer counts and distinct users, represented by \texttt{transfer\_count} and \texttt{daily\_users}.
It also includes value and fee aggregates when available, including \texttt{total\_amount\_usd}, \texttt{avg\_transfer\_usd\_value}, \texttt{total\_fee\_usd}, and \texttt{avg\_fee\_usd}.
Latency proxies are included when available, such as \texttt{avg\_speed\_seconds}.
We further report within-day distributional summaries, including \texttt{min}, \texttt{Q1}, \texttt{Q2}, \texttt{Q3}, and \texttt{max}, for transfer value, fees, and latency.

\subsection{Evolution of the Cross-Chain Corridor Network}

\begin{figure*}[t]
  \centering

  % 统一控制整张图的“列高度知道是多少”
  \newlength{\NetH}
  \setlength{\NetH}{0.46\textheight}

  % 每张小图的“图像高度”（注意：不含子标题）
  \newlength{\SmallImgH}
  \setlength{\SmallImgH}{0.285\NetH} % 三张合计约 0.855 NetH，剩下给 caption+间距

  % ---------- Left column: 3 small panels ----------
  \begin{minipage}[t][\NetH][t]{0.3\textwidth}
    \centering
    \vspace{0pt}

    \subcaptionbox{2022\label{fig:net-2022}}[\linewidth]{%
      \includegraphics[width=\linewidth,height=\SmallImgH,keepaspectratio]{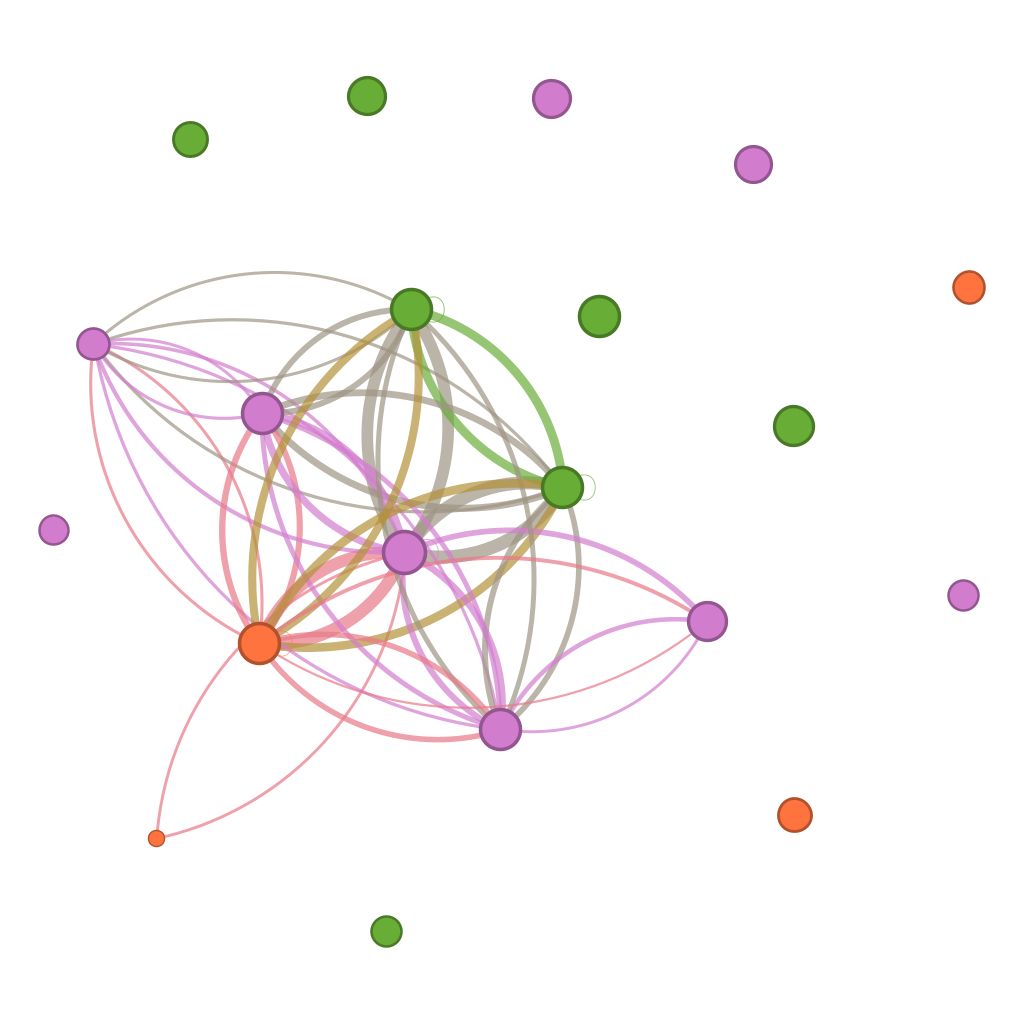}%
    }

    \vspace{3pt}

    \subcaptionbox{2023\label{fig:net-2023}}[\linewidth]{%
      \includegraphics[width=\linewidth,height=\SmallImgH,keepaspectratio]{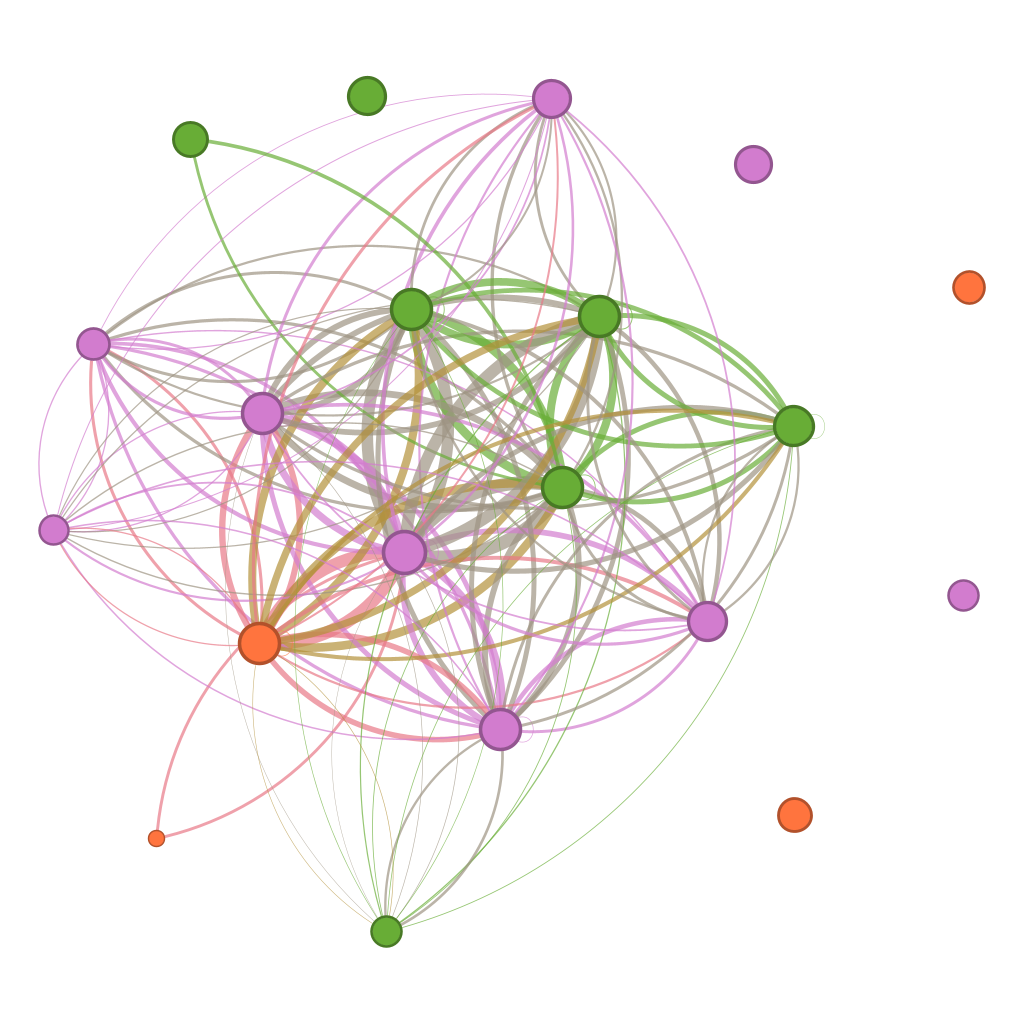}%
    }

    \vspace{3pt}

    \subcaptionbox{2024\label{fig:net-2024}}[\linewidth]{%
      \includegraphics[width=\linewidth,height=\SmallImgH,keepaspectratio]{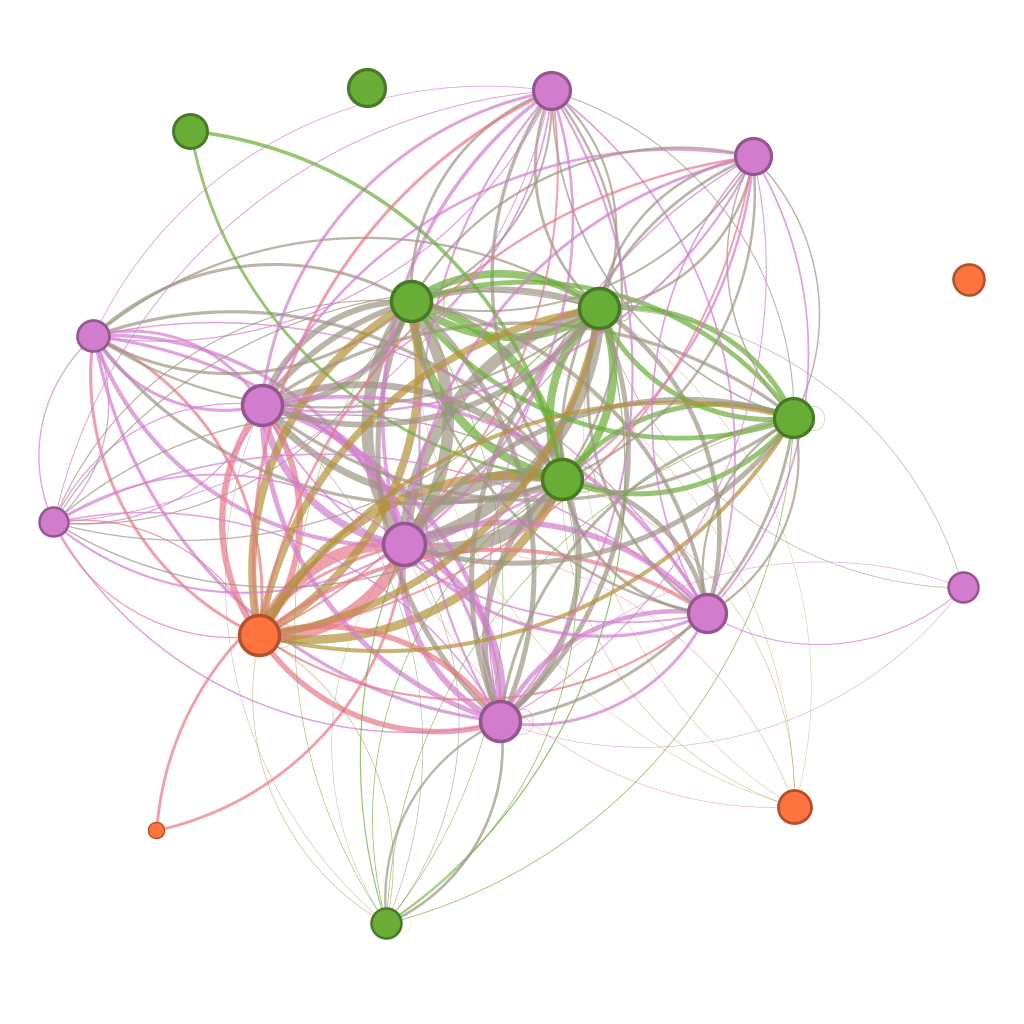}%
    }
  \end{minipage}
  \hfill
  % ---------- Right column: big panel ----------
  \begin{minipage}[t][\NetH][t]{0.68\textwidth}
    \centering
    \vspace{0pt}

    \subcaptionbox{2025 (labeled)\label{fig:net-2025}}[\linewidth]{%
      \includegraphics[width=\linewidth,height=0.93\NetH,keepaspectratio]{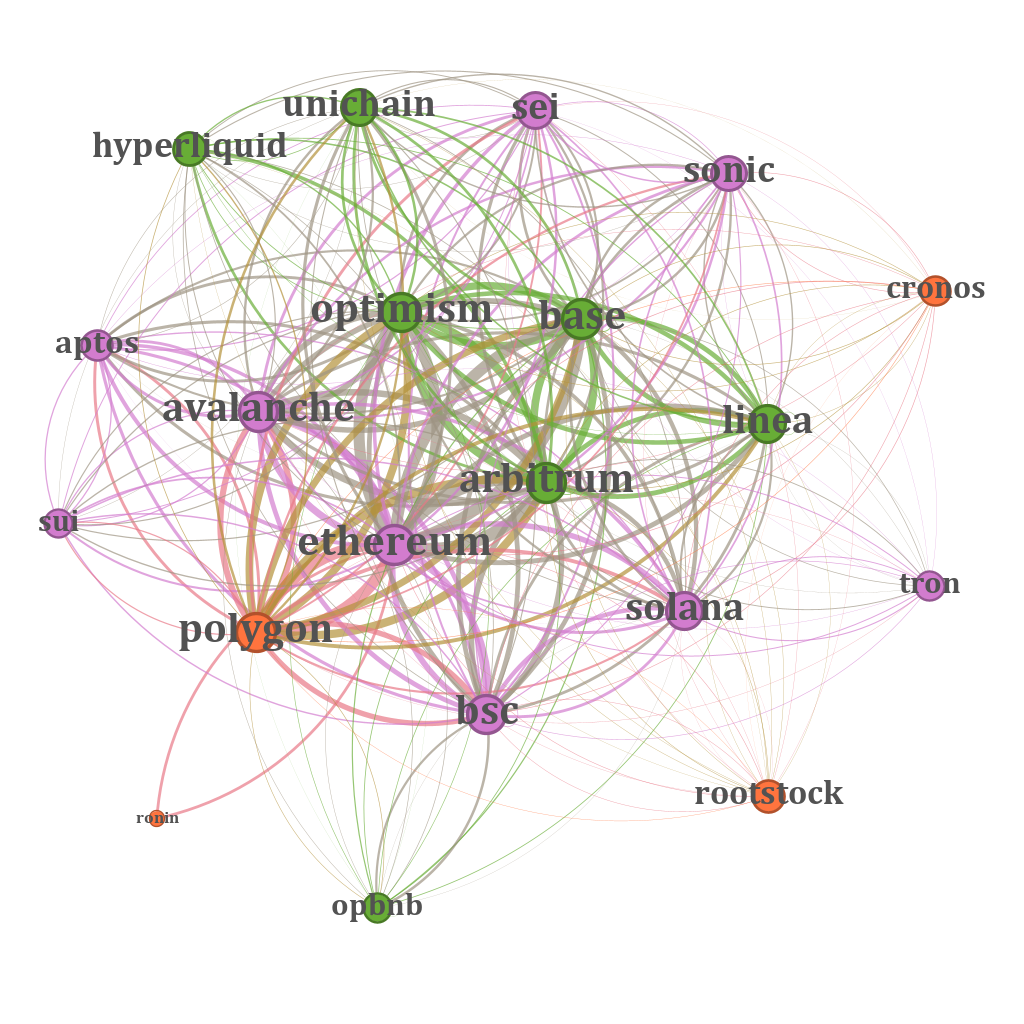}%
    }
  \end{minipage}

    \caption{Evolution of the cross-chain corridor network with annual snapshots from 2022 to 2025.
    Nodes are chains colored by stack type, with purple for Layer-1, green for Layer-2, and orange for sidechains.
    A directed edge indicates at least one observed transfer from a source chain to a destination chain during the year.
    Node size is proportional to total (in + out) degree in the annual corridor graph.
    Edge width is proportional to $\log_{10}(1 + V_{ij}^{y})$, where $V_{ij}^{y}$ is the total bridged USD notional on corridor $i\!\to\!j$ in year $y$.
    Node positions are fixed across years for comparability.}

  \label{fig:network_evolution}
\end{figure*}

Figure~\ref{fig:network_evolution} shows annual snapshots of the observed corridor network.
Nodes are chains and are colored by stack type.
Node size is proportional to total in-degree plus out-degree in the annual graph.
Edges are directed and bridge-specific.
An edge represents an active bridge--corridor, meaning that a given bridge carries transfers from a source chain to a destination chain within the year.
To keep panels comparable, we fix the node set and layout across years and update only the set of active edges and their weights.

Two patterns are salient.
First, the network densifies substantially from 2022 to 2025.
More corridors become active and the periphery attaches to a growing connected component.
Second, densification is not uniform.
By 2025 the ecosystem exhibits a clear multi-hub core dominated by large EVM-compatible Layer-1 and Layer-2 chains, including Ethereum, Arbitrum, Base, Optimism, Polygon, Avalanche, and Linea.
Several non-EVM or specialized chains remain more weakly connected and rely on fewer gateway corridors, including Solana, TRON, Cronos, Rootstock, Ronin, and opBNB.
Overall, interoperability expands in coverage, but activity remains concentrated on a subset of high-intensity corridors.
These topology changes motivate our later emphasis on chain integration and corridor coupling, rather than treating bridging as a single aggregate volume.

\subsection{Bridge-Level Activity and Cross-Ecosystem Corridors}

\begin{figure}[t]
    \centering
    \begin{subfigure}[t]{\linewidth}
        \centering
        \includegraphics[width=\linewidth]{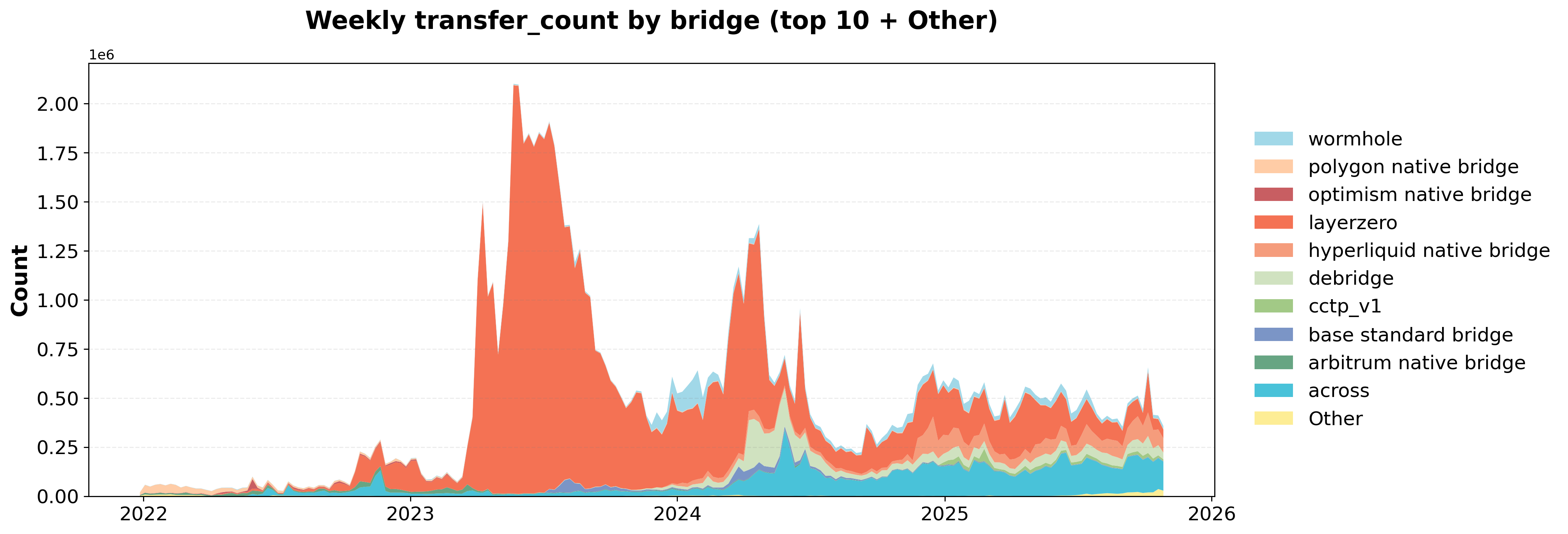}
        \caption{Weekly transfer counts.}
    \end{subfigure}

    \begin{subfigure}[t]{\linewidth}
        \centering
        \includegraphics[width=\linewidth]{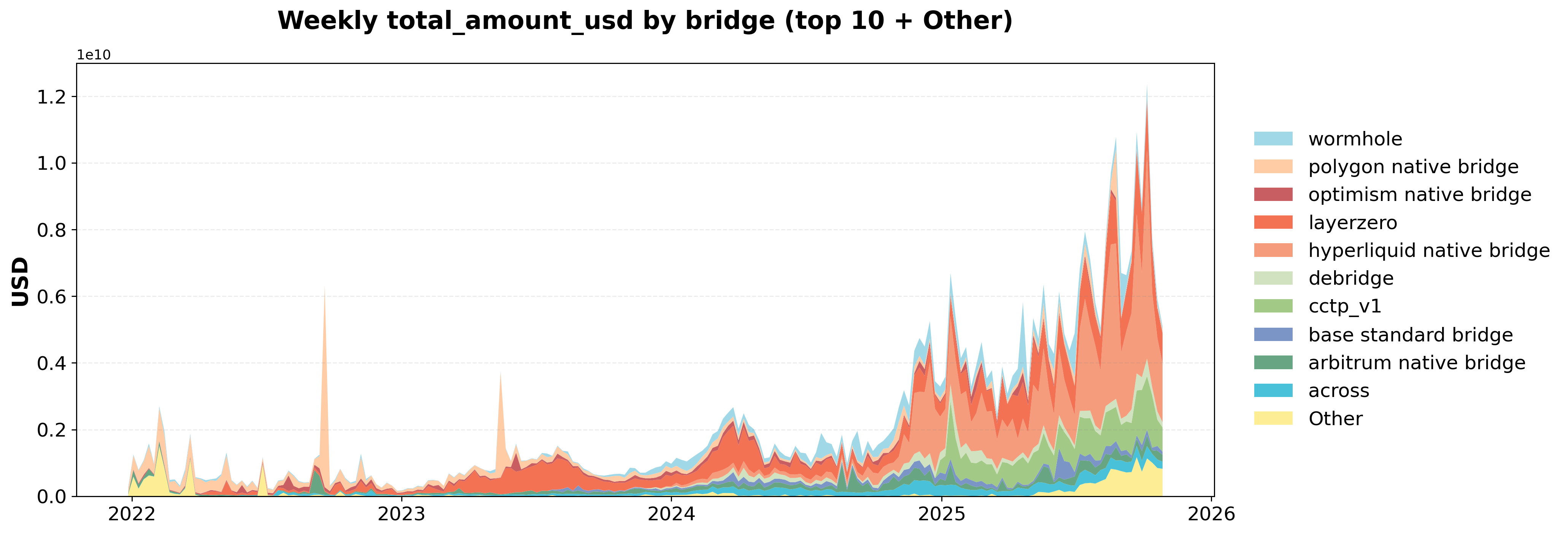}
        \caption{Weekly USD notional.}
    \end{subfigure}
    \caption{Weekly bridge-level activity (2022--2025). Stacked areas show the top-10 bridges ranked over the full window; remaining bridges are grouped as \textsc{Other}.}
    \label{fig:weekly_bridge_activity}
\end{figure}

\begin{figure}
    \centering
    \includegraphics[width=1\linewidth]{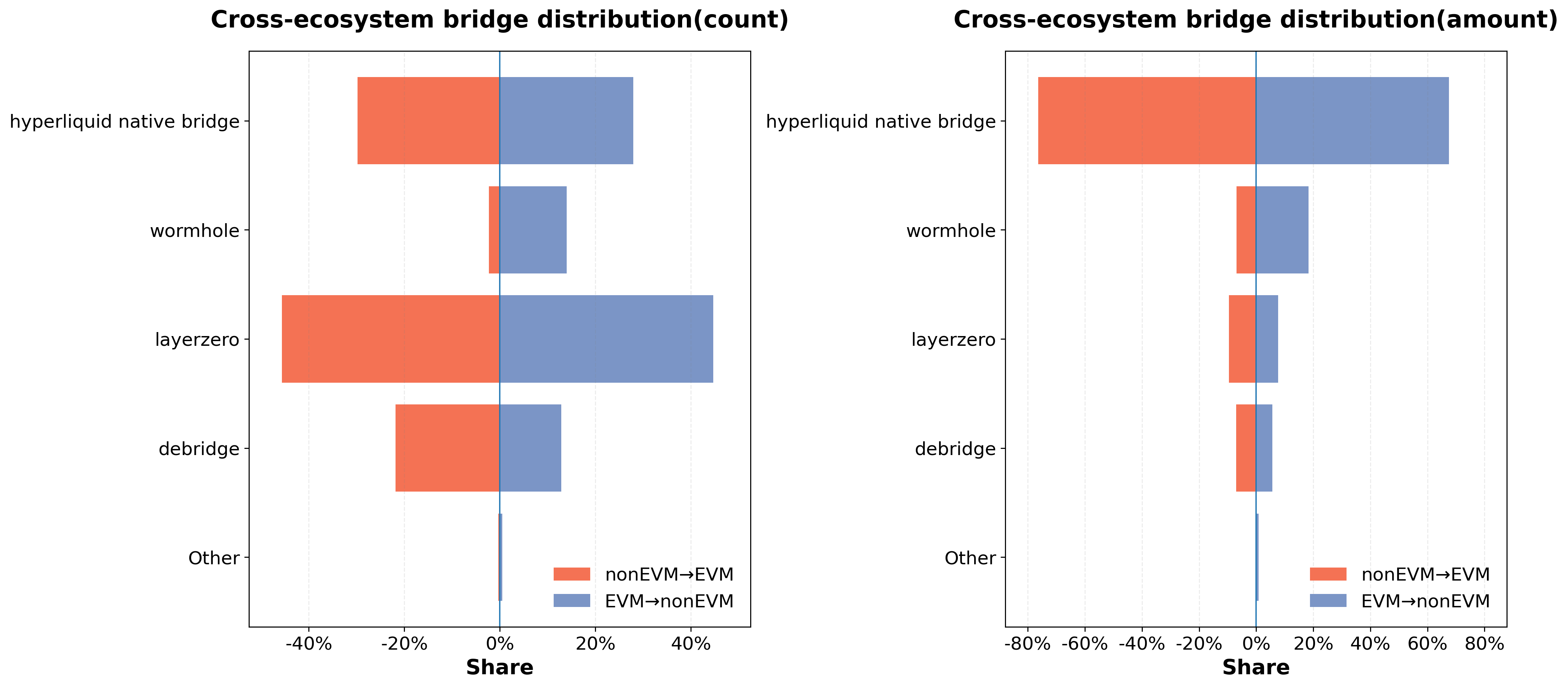}
    \caption{Bridge composition of cross-ecosystem traffic between EVM and non-EVM chains. Left: shares by weekly transfer counts; right: shares by USD notional. Shares are normalized within each direction (nonEVM$\rightarrow$EVM and EVM$\rightarrow$nonEVM). We show the top-4 bridges by aggregate cross-ecosystem activity; remaining bridges are grouped into \textsc{Other}.}

    \label{fig:evm_nonevm_split}
\end{figure}

Figure~\ref{fig:weekly_bridge_activity} summarizes weekly bridge-level activity from 2022 to 2025.
We plot the top-10 bridges by total activity over the full window and group the long tail into \textsc{Other}.
We report two complementary aggregates, transfer counts that capture usage intensity and USD notional that captures value mobility.
A key empirical regularity is that these two views highlight different episodes and different dominant bridges.

In counts, activity exhibits strong but transient concentration.
A prominent spike in 2023 coincides with widely discussed airdrop-related incentives around LayerZero-based routes, consistent with an incentive-driven surge in small, frequent transfers during that period~\cite{binance, dlnews}.
\footnote{LayerZero is an omnichain messaging layer that supports multiple bridge and application deployments. In our taxonomy, \textsc{LayerZero} aggregates several LayerZero-based routes, such as Stargate and USDT0, so the spike reflects an ecosystem-wide surge rather than a single-application anomaly.}
Afterward, the count composition becomes more diversified across multiple bridges.
In contrast, notional remains modest early on but rises sharply in late 2024 and 2025, with late-stage peaks concentrated in a small set of bridges.
Notably, the Hyperliquid native bridge accounts for a large share of high-notional weeks, illustrating that high-value interoperability can be dominated by chain-specific on and off ramps rather than the bridges that dominate low-value, high-frequency usage.

Figure~\ref{fig:evm_nonevm_split} links bridge composition to cross-ecosystem interoperability by grouping chains into EVM and non-EVM.
We report both directions, nonEVM$\rightarrow$EVM and EVM$\rightarrow$nonEVM, and normalize bridge shares within each direction.
Over the study window, cross-ecosystem totals are 6.36M versus 10.33M transfers and USD 68.3B versus 84.2B in notional for nonEVM$\rightarrow$EVM versus EVM$\rightarrow$nonEVM.
Two contrasts emerge.
First, by USD notional, late-stage peaks are concentrated in a small set of bridges, with the Hyperliquid native bridge accounting for a large share of high-notional weeks.
This pattern is consistent with the bridge acting as the primary funding and withdrawal rail for activity on Hyperliquid, such as USDC deposits to fund trading and withdrawals back to EVM addresses, which naturally concentrates large-value transfers.\footnote{Hyperliquid describes its bridge as a validator-signed mechanism with a dispute process; see \url{https://hyperliquid.gitbook.io/hyperliquid-docs/hypercore/bridge/}.}
Second, by counts, cross-ecosystem usage is dominated by LayerZero-based routes, consistent with application- and incentive-driven interactions that generate many small transfers.
Taken together, cross-ecosystem interoperability is asymmetric both in direction and in composition.
Bridges that dominate who interacts, measured by counts, need not dominate how much value moves, measured by notional.
For a global view of this count--notional asymmetry across bridges over the full window, see Appendix~\ref{app:bridge_count_amount_asym}.

\subsection{Chain-Level Participation and Net Flow Asymmetry}
\begin{figure}[t]
    \centering
    \begin{subfigure}[t]{\linewidth}
        \centering
        \includegraphics[width=\linewidth]{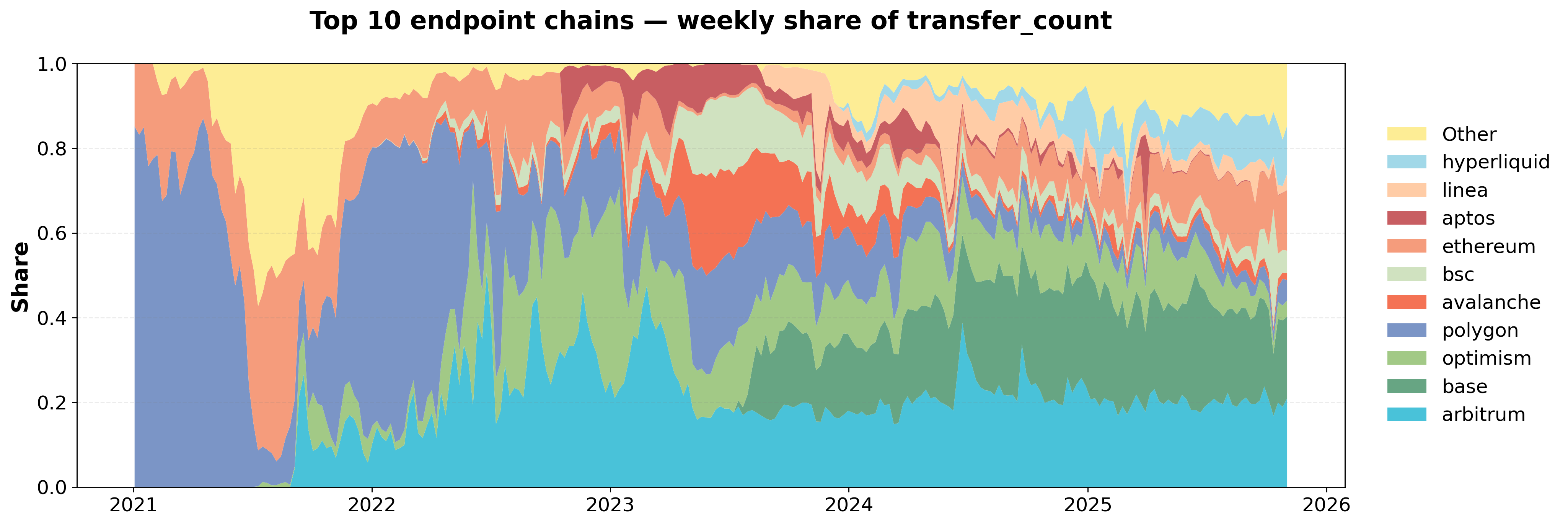}
        \caption{Endpoint share by transfer counts.}
        \label{fig:chain_weekly_count_share}
    \end{subfigure}

    \begin{subfigure}[t]{\linewidth}
        \centering
        \includegraphics[width=\linewidth]{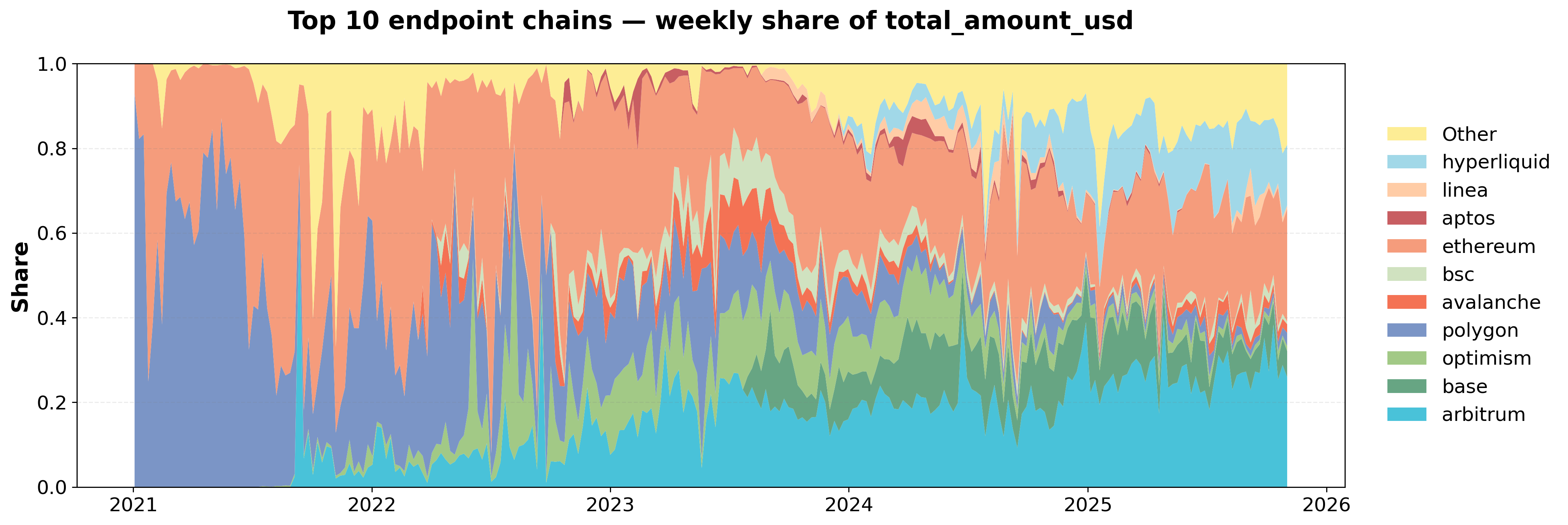}
        \caption{Endpoint share by USD notional.}
        \label{fig:chain_weekly_notional_share}
    \end{subfigure}

    \caption{Weekly endpoint share by chain (top-10 + \textsc{Other}). Each transfer is attributed to its endpoint chains and aggregated by week; stacked areas sum to one.}
    \label{fig:chain_weekly_share}
    \vspace{-0.8em}
\end{figure}

\begin{figure}
    \centering
    \includegraphics[width=0.9\linewidth]{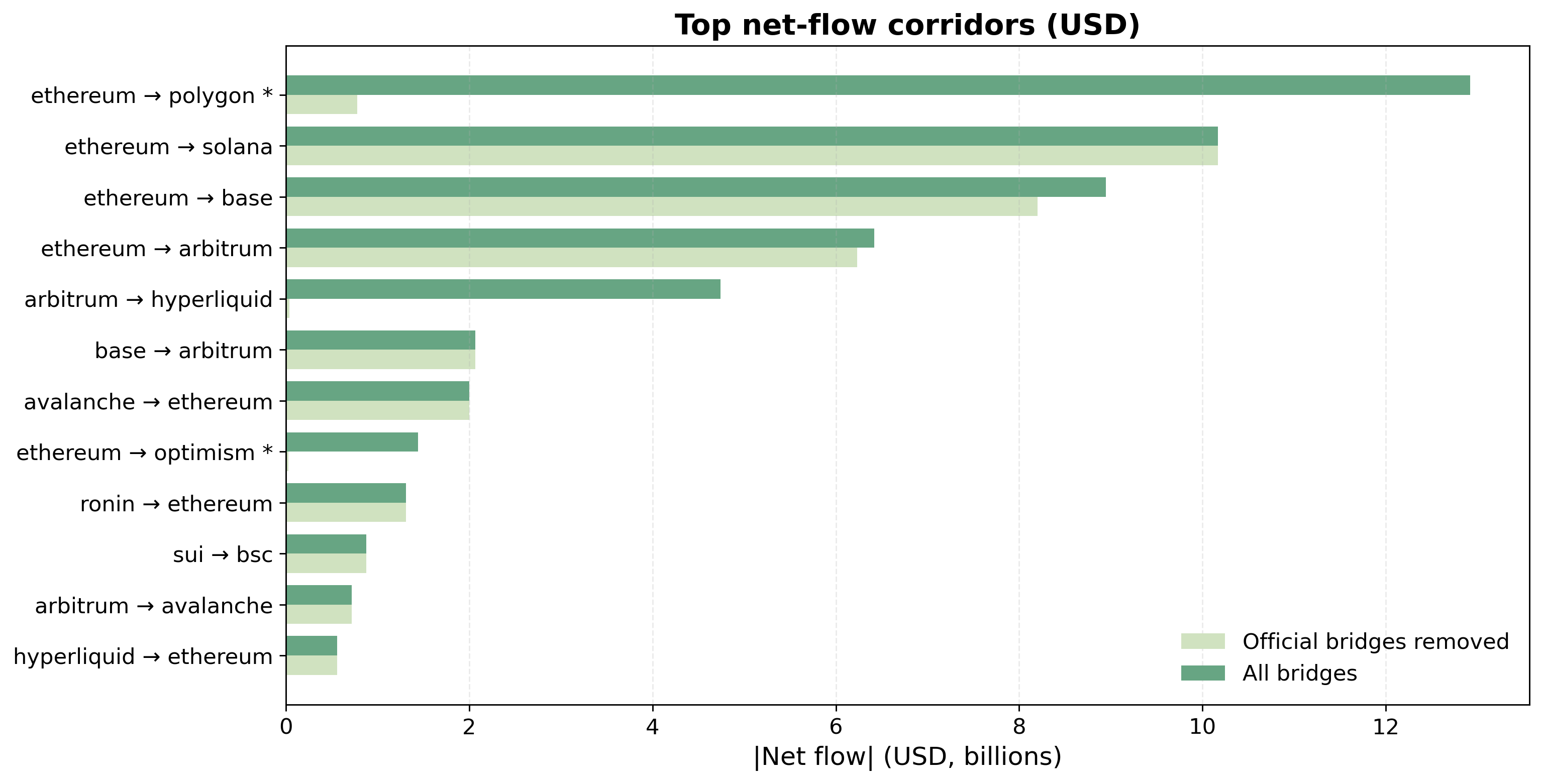}
    \caption{Largest absolute net flows by chain pair over the study window (USD billions). Arrows indicate net exporter $\rightarrow$ importer. Dark bars use all bridges; light bars exclude official/native bridges. (*) marks corridors whose net direction flips after exclusion.}

    \label{fig:net_flow_corridors}
\end{figure}
Figure~\ref{fig:chain_weekly_count_share} and Figure~\ref{fig:chain_weekly_notional_share} summarize weekly endpoint participation by chain.
We attribute each transfer to its endpoint chains, aggregate by week, and plot each chain's share among the top-10 endpoints, stacked to sum to one, separately for counts and notional.
The two series reveal persistent role heterogeneity.

On the count side, participation broadens over time.
Early activity is concentrated on a small set of endpoints, most notably Ethereum and Polygon.
Later periods show rising shares for major Layer-2 chains, including Arbitrum and Base, and the emergence of application-centric chains, such as Hyperliquid, as meaningful endpoints.
The rise of Base occurs shortly after the Dencun upgrade on March 13, 2024, consistent with lower Layer-2 operating costs enabling smaller and more frequent cross-chain interactions.\footnote{Dencun is the March 2024 Ethereum upgrade that activated EIP-4844, also known as proto-danksharding, and introduced blob-carrying transactions; see \url{https://ethereum.org/roadmap/dencun/} and EIP-4844~\cite{eip4844}.}
On the notional side, concentration remains stronger.
Ethereum remains a major value endpoint throughout much of the window, but its share declines in later periods as Arbitrum and Hyperliquid account for a larger fraction of cross-chain value.
Several chains contribute substantial endpoint counts but comparatively small notional, indicating that popular endpoints by frequency are not necessarily major endpoints by value.

Figure~\ref{fig:net_flow_corridors} further characterizes directional roles using net flows aggregated over all bridges.
A small set of corridors accounts for the largest absolute net flows, anchored around Ethereum and a few large Layer-2 and application-centric ecosystems.
Excluding official and native bridges, shown by light bars, changes the magnitude and, for some corridors, even the net direction, marked by $^\ast$.
This highlights that canonical routes can materially shape measured net positions.
These directional asymmetries motivate our corridor-level metrics and the subsequent analyses of how structural connectivity and realized flows relate to congestion and risk coupling.

% % % % % % % % % % % % % % % % % % % % % % % % % % % % % % % % % % % % % % % % % % % % 
\section{The Causal Effects of Interoperability: Paradox and Trade-offs}
\label{sec:causal}

In this section, we investigate the causal effects of interoperability on the blockchain economy. We progress from identifying the drivers of capital flow to revealing two systemic phenomena: the \emph{Growth-Return Paradox} and the \emph{Efficiency-Fragility Trade-off}.

\subsection{Empirical Strategy and Identification}

Our main analyses quantify how interoperability relates to (i) future net capital inflows, (ii) chain-level economic activity, and (iii) congestion and risk contagion, using fixed-effect panel regressions and a difference-in-differences (DiD) design.
These tools separate persistent cross-chain differences from short-run changes and strengthen causal interpretation under large common market shocks.
The unit of observation is chain-day, except for contagion regressions where it is chain-pair-day.

\paragraph{Panel regressions with fixed effects.}
Most analyses use the following chain-day panel specification:
\begin{equation}
Y_{i,t} = \alpha + \beta_1 \text{ASI}_{i,t} + \beta_2 \text{AAI}_{i,t} + \gamma^\top X_{i,t} + \mu_i + \lambda_t + \varepsilon_{i,t},
\end{equation}
where $Y_{i,t}$ is a chain-level outcome, $X_{i,t}$ includes control variables such as recent returns and gas metrics, $\mu_i$ are chain fixed effects, and $\lambda_t$ are day fixed effects.
Chain fixed effects remove time-invariant properties of each blockchain, such as architecture, developer culture, or baseline user base. Day fixed effects remove market-wide shocks that affect all chains simultaneously, such as bull or bear cycles. Therefore, coefficients are identified from deviations within the same chain relative to its own typical level, on days when overall market conditions are similar. This helps distinguish structural associations from simple size effects or global trends.
This framework is used to study:
(i) drivers of net capital inflows,
(ii) effects on TVL, users, and developer activity,
(iii) effects on token returns, and
(iv) congestion and transaction costs.

\paragraph{Pairwise comovement regressions.}
To study contagion, we analyze chain-pair synchronization using rolling-window correlations of TVL:
\begin{equation}
\rho^{(w)}_{ij,t} = \alpha + \theta_1 \text{PSI}_{ij,t} + \theta_2 \text{Flow}_{ij,t} + \mu_{ij} + \lambda_t + u_{ij,t},
\end{equation}
where $\mu_{ij}$ are pair fixed effects. This design isolates how changes in structural or active interoperability between the two chains are associated with changes in their economic synchronization over time.

\paragraph{Difference-in-differences using bridge failure.}
To strengthen causal interpretation, we exploit the collapse of the Multichain bridge on July 6th, 2023 as an exogenous disruption to cross-chain trust:
\begin{equation}
Y_{i,t} = \alpha + \delta (\text{Treat}_i \times \text{Post}_t) + \gamma^\top X_{i,t} + \mu_i + \lambda_t + \epsilon_{i,t}.
\end{equation}
Here, treated chains are those supported by Multichain. The interaction coefficient $\delta$ measures whether these chains experienced a differential post-shock decline relative to other chains, after controlling for time trends and chain-specific baselines. The DiD design leverages a discrete external event and thus provides stronger evidence of causal impact from interoperability-related failures.

\subsection{Capital Flow Drivers: Incentives vs. Infrastructure}

Understanding what drives cross-chain capital movements is central for interpreting interoperability.
We test whether capital flows are driven by infrastructure availability or economic incentives. We find that flows are return-chasing and cost-sensitive, while structural interoperability acts as a neutral enabler rather than a directional driver.

Table \ref{tab:drivers_capital_flow} examines the determinants of future 7-day net capital inflow. The models evaluate the predictive power of current token returns, economic variables, and gas cost metrics on future capital direction, measured at day $t$.
Three insights stand out: (i) return-chasing: net inflows are strongly associated with recent native-token performance. The \emph{Token Return} coefficient is significantly positive, and remains robust with chain and day fixed effects. This indicates that cross-chain capital tends to move toward ecosystems that have recently delivered higher returns. (ii) cost-sensitive: transaction-cost frictions matter at the margin. \emph{Avg Gas/Tx} is negative and statistically significant, implying higher per-transaction costs predict weaker future net inflows. Economically, this aligns with the idea that cross-chain reallocations are not frictionless: even if users can bridge, they still prefer lower execution costs when deciding where to deploy capital.
(iii) the neutrality of interoperability: \emph{ASI} is small and statistically indistinguishable from zero, suggesting infrastructure capacity does not determine flow direction; instead, direction is shaped by contemporaneous incentives.

\begin{table}[htbp]
  \centering
  \begin{threeparttable}
    \caption{Drivers of Net Capital Inflow}
    \label{tab:drivers_capital_flow}
    \footnotesize 
    \setlength{\tabcolsep}{1.5pt}
    
    \begin{tabular}{l c c c c c c c c c c}
      \toprule
       & Return & TVL & DAU & New Contr. & ASI & Gas/Tx & Gas Fee & Gas Used & Obs & $R^2$ \\
      \midrule
      (1) & \textbf{1.061***} & -0.017 & -0.004 & -0.012 & 0.002 & & & & 17,409 & 0.069 \\
          & (5.16) & (-1.18) & (-0.24) & (-0.67) & (0.92) & & & & & \\
      \addlinespace
      (2) & \textbf{1.038***} & 0.011 & -0.041 & -0.000 & 0.004 & \textbf{-0.061**} & 0.010 & -0.063 & 13,349 & 0.090 \\
          & (4.06) & (0.54) & (-0.62) & (-0.02) & (0.56) & (-2.45) & (0.62) & (-1.09) & & \\
      \midrule
      \multicolumn{11}{l}{Dep. Var: Net Inflow; Day FE: Yes; Chain FE: Yes} \\
      \bottomrule
    \end{tabular}
    
    \begin{tablenotes}[flushleft]
      \tiny
      \setlength{\itemindent}{-2.5pt}
      \item \textit{Notes:} The dependent variable is the Future 7-Day Net Inflow (Signed Log). 
      Predictors are current day values.
      $t$-statistics in parentheses. * p$<$0.1, ** p$<$0.05, *** p$<$0.01
    \end{tablenotes}
  \end{threeparttable}
\end{table}

\subsection{The Growth-Return Paradox: Ecosystem Expansion vs. Token Dilution} 

Does higher interoperability create value?
We uncover a \emph{Growth-Return Paradox}: while connectivity expands on-chain ecosystem scale, it exerts downward pressure on native token returns due to supply-side composition effects.

\paragraph{Interoperability and Ecosystem Scale}
We find that both types of interoperability can improve TVL, participation, and contract deployment.

Table~\ref{tab:econ_growth} therefore estimates fixed-effect panel regressions. Panel A shows that ASI is positively and precisely associated with TVL, DAU (daily active users), and new contract deployments. The coefficients are statistically significant across all three outcomes, and the pattern is consistent with a network-access channel: as a chain becomes better connected to economically deep neighbors, its local ecosystem tends to expand in size and participation, with a parallel increase in development activity.
Panel B relates outcomes to AAI, which measures realized cross-chain utilization. The coefficients are also statistically significant across outcomes, and their magnitudes are economically meaningful. Active interoperability mechanically reflects realized interaction intensity, which tends to co-move with on-chain usage and liquidity deployment. 

\begin{table}[htbp]
  \centering
  \begin{threeparttable}
    \caption{Impact of ASI and AAI on Economic Growth}
    \label{tab:econ_growth}
    \footnotesize 
    \setlength{\tabcolsep}{5pt}
    \begin{tabular}{l c c c}
      \toprule
      & (1) & (2) & (3) \\
      & TVL (MA7) & DAU (MA7) & New Contracts (MA7) \\
      \midrule
      
      % --- Panel A ---
      \multicolumn{4}{l}{\textbf{Panel A: ASI}} \\
      \midrule
        ASI & \textbf{0.032***} & \textbf{0.019***} & \textbf{0.043***} \\
          & (16.21) & (10.49) & (23.39) \\
      \addlinespace
      Observations & 17,399 & 17,399 & 17,399 \\
        $R^2$ & 0.481 & 0.506 & 0.254 \\
      
      \midrule
      
      % --- Panel B ---
      \multicolumn{4}{l}{\textbf{Panel B: AAI}} \\
      \midrule
       AAI & \textbf{0.742***} & \textbf{2.220***} & \textbf{0.419***} \\
          & (31.30) & (18.28) & (3.24) \\
      \addlinespace
      Observations & 17,399 & 17,399 & 17,399 \\
      $R^2$ & 0.984 & 0.513 & 0.189 \\
      \midrule
      \multicolumn{4}{l}{Controls: Yes; Chain FE: Yes; Day FE: Yes} \\
      \bottomrule
    \end{tabular}
    
\begin{tablenotes}[flushleft]
      \tiny
      \setlength{\itemindent}{-2.5pt}
      \item \textit{Notes:} This table reports the regression results for the impact of Aggregated Structural Interoperability (ASI) and Aggregated Active Interoperability (AAI). 
      Panel A presents the estimates for the ASI index, while Panel B presents the estimates for the AAI index. 
      The dependent variables are Total Value Locked (TVL) in column (1),  Daily Active Users (DAU) in column (2), and New Contracts in column (3). 
      All regressions control economic and cost variables, except themselves. 
      Chain fixed effects (Chain FE) and Day fixed effects (Day FE) are included in all specifications. 
      $t$-statistics are reported in parentheses. 
      *, **, and *** denote statistical significance at the 10\%, 5\%, and 1\% levels, respectively.
    \end{tablenotes}
  \end{threeparttable}
\end{table}

\paragraph{Interoperability and Token Returns}
While activity expands, Table~\ref{tab:return_term_structure} shows that interoperability is negatively associated with native token returns.
ASI shows no short-term effect but becomes significantly negative at longer horizons. AAI is significantly negative at medium and long horizons.

We propose a ``Supply-Side Composition Hypothesis'' to explain these counter-intuitive patterns: higher ASI can lower native token prices by shifting on-chain collateral composition.
\begin{itemize}
    \item CEX vs.\ on-chain routing: when ASI is low and cross-chain friction is high, users rely more on CEXs; CEX routing is an off-chain ledger swap that reallocates ownership without changing aggregate on-chain asset supplies.
    \item Hard-asset outflow under bridging: as ASI increases, users shift toward on-chain bridging. Transfers are often denominated in hard assets, e.g., wBTC, wETH; bridging these out reduces the local supply of strong collateral assets.
    \item Relative native-token oversupply: while ASI facilitates hard-asset outflows, the native token supply is relatively stable (or increases via incentives), reducing the hard-asset-to-native ratio; the resulting relative oversupply can depress prices and future returns, consistent with the negative coefficients.
\end{itemize}

\begin{table}[htbp]
  \centering
  \begin{threeparttable}
    \caption{Term Structure of Impact on Native Token Returns}
    \label{tab:return_term_structure}
        \footnotesize 
    \setlength{\tabcolsep}{5pt}
    \begin{tabular}{l c c c c c c}
      \toprule
      & (1) & (2) & (3) & (4) & (5) & (6) \\
      & 1-Day & 3-Day & 7-Day & 15-Day & 30-Day & 100-Day \\
      \midrule
    ASI & -0.000 & -0.000 & \textbf{-0.001**} & \textbf{-0.002***} & \textbf{-0.004***} & \textbf{-0.011***} \\
          & (-0.67) & (-1.28) & (-2.15) & (-3.26) & (-4.12) & (-5.93) \\
      \addlinespace
      AAI & -0.011 & \textbf{-0.035*} & \textbf{-0.076**} & \textbf{-0.128***} & \textbf{-0.139**} & \textbf{-0.294**} \\
          & (-0.91) & (-1.67) & (-2.45) & (-2.87) & (-2.14) & (-2.43) \\
      \midrule
      \multicolumn{7}{l}{Controls: Yes; Chain FE: Yes; Day FE: Yes} \\
      \midrule
        Observations & 17,363 & 17,326 & 17,258 & 17,122 & 16,867 & 15,682 \\
        $R^2$ & 0.135 & 0.141 & 0.143 & 0.149 & 0.150 & 0.135 \\
      \bottomrule
    \end{tabular}
    
    \begin{tablenotes}[flushleft]
      \tiny
      \setlength{\itemindent}{-2.5pt}
      \item \textit{Notes:} The dependent variables are cumulative returns over future $k$ days (where $k$ = 1, 3, 7, 15, 30, 100). 
      Sample size decreases for longer horizons due to the forward-looking calculation window.
      All models include control variables, Chain FE, and Day FE.
      $t$-statistics are reported in parentheses.
      * p$<$0.1, ** p$<$0.05, *** p$<$0.01
    \end{tablenotes}
  \end{threeparttable}
\end{table}

\subsection{Systemic Trade-offs: Decongestion Efficiency vs. Contagion Fragility}

Then, we analyze the externalities of interoperability. We identify a fundamental trade-off: structural interoperability is associated with lower congestion costs, while active interoperability raises costs and strengthens cross-chain coupling.
At the same time, higher interoperability increases fragility by amplifying spillovers from bridge failure and synchronized chains.

\subsubsection{Efficiency: Decongestion via Load Balancing.}

Interoperability can both enable load balancing and create additional demand.
We document that structural interoperability can lower the average transaction cost, while active interoperability adds congestion.

Table~\ref{tab:congestion_costs} reveals the sharp contrast. ASI is negatively associated with gas fees and usage, consistent with capacity enabling offloading to alternative execution layers.
Conversely, AAI is positively associated with congestion and costs. Realized transfers generate execution demand that competes for block space. Thus, capacity facilitates load balancing, while utilization consumes resources.

\begin{table}[htbp]
  \centering
  \begin{threeparttable}
    \caption{Impact of Interoperability on Congestion and Costs}
    \label{tab:congestion_costs}
        \footnotesize 
    \setlength{\tabcolsep}{10pt}
    \begin{tabular}{l c c c}
      \toprule
      & (1) & (2) & (3) \\
      & Gas Fee & Gas Used & Avg Gas/Tx \\
      \midrule
      ASI & \textbf{-0.053***} & \textbf{-0.033***} & \textbf{-0.007***} \\
          & (-7.37) & (-5.79) & (-2.88) \\
      \addlinespace
      AAI & \textbf{3.887***} & \textbf{0.813**} & \textbf{0.886***} \\
          & (7.91) & (2.11) & (8.51) \\
      \midrule
      \multicolumn{4}{l}{Controls: Yes; Chain FE: Yes; Day FE: Yes} \\
      \midrule
      Observations & 17,402 & 17,402 & 13,373 \\
      $R^2$ & 0.151 & 0.335 & 0.163 \\
      \bottomrule
    \end{tabular}
    
    \begin{tablenotes}[flushleft]
      \tiny
      \setlength{\itemindent}{-2.5pt}
      \item \textit{Notes:} The table reports the impact of ASI and AAI on network congestion and costs. 
      Dependent variables are Total Gas Fee (Column 1), Total Gas Used (Column 2), and Average Gas Price per Transaction (Column 3).
      $t$-statistics are reported in parentheses.
      * p$<$0.1, ** p$<$0.05, *** p$<$0.01
    \end{tablenotes}
  \end{threeparttable}
\end{table}

\subsubsection{Fragility: Attack Risk Exposure and Contagion.}
Efficiency gains come with heightened systemic risk. We assess this through direct exposure to bridge failures and synchronization of economic shocks.

\paragraph{Risk Exposure from Bridge Failures}

Tight coupling can reduce failure isolation and increase systemic fragility; bridges create such coupling across blockchains. If shocks propagate more easily under higher connectivity, bridge infrastructure becomes a contagion channel, which motivates tests based on economic synchronization.

We use the Multichain bridge collapse as an exogenous shock in a DiD framework. Table \ref{tab:didmultichain} shows that treated chains, those dependent on Multichain, experienced a sharp, causal decline in TVL relative to control chains. The interaction term is negative and significant ($-0.772$, $p<0.01$), indicating that connectivity acts as a direct vector for shock transmission.
These results highlight a vulnerability of interoperability: it supports capital efficiency in normal times but amplifies tail risk. A central bridge failure can transmit systemic shocks through wrapped-asset de-pegging, generating asymmetric damage across connected ecosystems, consistent with contagion risk from connectivity.

\begin{table}[htbp]
  \centering
  \begin{threeparttable}
    \caption{Impact of Multichain Collapse on Chain Economy}
    \label{tab:didmultichain}
    \small
    \setlength{\tabcolsep}{6pt}
    \begin{tabular}{l c c c c c c}
      \toprule
      Dep. Var & Treat$\times$Post & Controls & Chain FE & Day FE & Obs & $R^2$ \\
      \midrule
      \addlinespace
      TVL & \textbf{-0.772***} & Yes & Yes & Yes & 17,399 & 0.468 \\
      \addlinespace
      \bottomrule
    \end{tabular}
    
    \begin{tablenotes}[flushleft]
      \scriptsize
      \setlength{\itemindent}{-2.5pt}
      \item \textit{Notes:} This table reports the difference-in-differences estimate around the Multichain collapse (2023-07-06).
      * p$<$0.1, ** p$<$0.05, *** p$<$0.01
    \end{tablenotes}
  \end{threeparttable}
\end{table}

\paragraph{Risk Contagion and Pairwise Synchronization}
We next study pairwise mechanisms using TVL correlations between chains $i$ and $j$, $\rho_{TVL}$, as a proxy for economic synchronization and potential risk transmission.
We compute rolling correlations over 30-, 60-, and 90-day windows; higher $\rho_{TVL}$ implies shocks in one chain are more likely to be mirrored in the other.

Table \ref{tab:contagion} shows that both structural connectivity and capital flows increase synchronization.
PSI has a positive and significant effect that grows with the window (0.004 at 30D to 0.033 at 90D), indicating infrastructure links synchronize cycles over longer horizons.
Total flow is also positive and highly significant (about 0.007--0.009), consistent with active channels serving as vectors of transmission.
Together, these results identify a negative externality: connectivity improves capital mobility but builds a tighter network for contagion, making idiosyncratic shocks more likely to become system-wide fluctuations.

\begin{table}[htbp]
  \centering
  \begin{threeparttable}
    \caption{Impact of Interoperability on TVL Correlation}
    \label{tab:contagion}
        \footnotesize 
    \setlength{\tabcolsep}{5pt}
    \begin{tabular}{l c c c}
      \toprule
      & (1) & (2) & (3) \\
      & $\rho_{TVL}$ (30D) & $\rho_{TVL}$ (60D) & $\rho_{TVL}$ (90D) \\
      \midrule
      PSI & \textbf{0.004**} & \textbf{0.025***} & \textbf{0.033***} \\
          & (2.28) & (17.06) & (24.59) \\
      \addlinespace
      Total flow & \textbf{0.007***} & \textbf{0.009***} & \textbf{0.008***} \\
          & (23.27) & (34.60) & (35.67) \\
      \midrule
      Observations & 85,506 & 85,506 & 85,506 \\
      $R^2$ & 0.354 & 0.386 & 0.404 \\
      \bottomrule
    \end{tabular}
    
    \begin{tablenotes}[flushleft]
      \tiny
      \setlength{\itemindent}{-2.5pt}
      \item \textit{Notes:} $\rho_{TVL}$ is the correlation between the TVL of the pair of chains, calculated in the rolling window of 30 days, 60 days, and 90 days.
      $t$-statistics are reported in parentheses.
      * p$<$0.1, ** p$<$0.05, *** p$<$0.01
    \end{tablenotes}
  \end{threeparttable}
\end{table}

% % % % % % % % % % % % % % % % % % % % % % % % % % % % % % % % % % % % % % % % % % % % 
\section{Architectural Heterogeneity: Why Design Matters}
\label{sec:architecture}

This section examines how the aggregate effects identified in Section \ref{sec:causal} vary across architectural settings. We show that heterogeneity is driven by (i) chain architecture and (ii) bridge architecture.
The key finding is that interoperability benefits are strongest when supported by execution compatibility and specific bridge designs.

\subsection{Chain Architecture: Execution Compatibility and Layer Roles}
Differences in virtual machine compatibility determine whether workloads can migrate seamlessly between chains. We find that EVM compatibility is a prerequisite for vertical scaling, whereas non-EVM chains suffer from liquidity extraction. 

We define \texttt{isEVM} for EVM-compatible chains and \texttt{isL1} for Layer~1 chains, and interact ASI/AAI with these indicators.
With interactions, the main ASI (AAI) coefficient applies to chains that are neither EVM-compatible nor L1, and interaction terms capture differences for EVM and for L1.
Table~\ref{tab:heterogeneity_matrix} reports implied effects for four groups (EVM vs.\ non-EVM, L1 vs.\ L2); full regressions are in Appendix Table~\ref{tab:heterogeneity_asi_aai}. 

Panel A shows that for EVM chains, ASI is positively associated with TVL on both L1 and L2. Crucially, ASI lowers gas costs on L1 while increasing gas usage on L2. This pattern is consistent with load shifting within a standardized stack. Because contracts remain compatible, connectivity allows users to offload execution to L2s while retaining shared liquidity, realizing the efficiency benefits of interoperability.
In contrast, for non-EVM chains, ASI increases transaction costs without boosting TVL. 

Panel B shows that AAI is strongly negatively associated with TVL.
Without execution compatibility, bridges cannot migrate workloads. Instead, they act as vectors for vampire attacks, facilitating capital flight to other ecosystems and leaving the local chain with congestion but without growth.

\begin{table}[htbp]
  \centering
  \begin{threeparttable}
    \caption{Heterogeneity Analysis by Chain EVM Compatibility and Layer}
    \label{tab:heterogeneity_matrix}
        \footnotesize 
    \setlength{\tabcolsep}{10pt}
    \begin{tabular}{l c c c c}
      \toprule
      & \multicolumn{2}{c}{TVL} & \multicolumn{2}{c}{Avg Gas/Tx} \\
      \cmidrule(lr){2-3} \cmidrule(lr){4-5}
      & L1 & L2 & L1 & L2 \\
      \midrule
      \multicolumn{5}{l}{\textbf{Panel A: ASI}} \\
       \midrule
      EVM & \textbf{0.017***} & \textbf{0.031***} & \textbf{-0.003***} & \textbf{0.003***} \\
      Non-EVM & \textbf{-0.017***} & -0.003 & \textbf{0.014***} & \textbf{0.020***} \\
      \midrule
      \multicolumn{5}{l}{\textbf{Panel B: AAI}} \\
       \midrule
      EVM & \textbf{4.336***} & \textbf{-2.827***} & \textbf{4.533***} & \textbf{0.990***} \\
      Non-EVM & \textbf{-2.232***} & \textbf{-9.395***} & \textbf{-0.998***} & \textbf{-4.541***} \\
      \bottomrule
    \end{tabular}
    
    \begin{tablenotes}[flushleft]
      \tiny
      \setlength{\itemindent}{-2.5pt}
      \item \textit{Notes:} This table reports the coefficients of ASI (Panel A) and AAI (Panel B) across different types of chain.
      Columns are split by Layer-1 (L1) vs Layer-2 (L2) status. Rows are split by EVM compatibility. See Appendix for the complete regression result.
      *** p$<$0.01
    \end{tablenotes}
  \end{threeparttable}
\end{table}

\subsection{Bridge Architecture: Asset Transfer Mechanisms and Trust Patterns} 

A natural next question is whether these effects differ systematically across bridge designs. 
We decompose interoperability by trust patterns and transfer mechanisms to understand how specific bridge design shapes economic outcomes.

\subsubsection{Official versus third-party bridges}
Distinguishing between official and third-party bridges reveals whether interoperability serves ecosystem integration or external liquidity acquisition.
Table~\ref{tab:heterogeneity_bridge_type} splits interoperability into contributions from official bridges versus third-party bridges.

Panel A indicates that \emph{official bridge ASI} is negatively associated with chain TVL and lower gas costs. This aligns with the function of vertical migration (e.g., L1 to L2). Official bridges relieve L1 congestion by moving assets to L2, which may reduce TVL but improves system efficiency.
In contrast, \emph{third-party bridge ASI} is positively associated with TVL. This aligns with horizontal growth. Third-party bridges primarily serve to acquire external users and liquidity from competing ecosystems.

Panel B shows that active interoperability has different signs across bridge types for TVL:  \emph{official bridge AAI} is strongly negatively associated with TVL, while \emph{third-party bridge AAI} is positively associated with TVL. Both types are positively associated with higher average gas cost per transaction, consistent with the idea that realized transfers generate additional execution demand regardless of verification model. Taken together, the type split suggests that bridge design affects not only security assumptions, but also which economic margin the bridge primarily serves: canonical routes appear more aligned with execution and migration, while third-party routes appear more aligned with liquidity sourcing and cross-ecosystem composition.

\subsubsection{Transfer Mechanisms: Lock-and-Mint, Burn-and-Mint, and Liquidity Pools}
Asset transfer mechanisms dictate on-chain state requirements. 
Table~\ref{tab:heterogeneity_mechanism} further splits interoperability by asset-transfer mechanism.

Panel A shows structural interoperability via lock-and-mint and liquidity pools is positively associated with TVL, whereas burn-and-mint is negatively associated. A plausible explanation is that burn-and-mint designs are concentrated in issuer-controlled token protocols, e.g., USDC, where interoperability increases mobility of a specific asset but need not increase aggregate TVL.
Liquidity-pool designs can attract capital to provision pools and support routing, linking more directly to TVL.
Regarding congestion, liquidity pool interoperability is associated with lower gas costs.
Liquidity networks can internalize routing and alignment within pools and reduce repeated actions in some settings, whereas lock-and-mint routes induce explicit endpoint settlement actions.

Panel B, however, reveals the cost of state. AAI strongly increases gas costs across all types, but liquidity pools show the largest magnitude. This highlights a scalability trade-off. While liquidity pools optimize routing structurally, their active rebalancing operations impose the highest execution overhead, whereas stateless mechanisms, like burn-and-mint, scale better with volume.

\begin{table}[htbp]
  \centering
  \begin{minipage}[t]{0.49\textwidth}
    \centering
    \begin{threeparttable}
      \caption{Heterogeneity Analysis by Bridge Type}
      \label{tab:heterogeneity_bridge_type}
      \scriptsize
      \setlength{\tabcolsep}{2pt} 
      \begin{tabular}{l c c c c}
        \toprule
        Dep. Var & $A*I_{Offi}$ & $A*I_{TP}$ & Obs & $R^2$ \\
        \midrule
        \multicolumn{5}{l}{\textbf{Panel A: ASI}} \\
        \addlinespace
        TVL & \textbf{-0.273***} & \textbf{0.060***} & 17,399 & 0.500 \\
            & (-20.08) & (26.52) & & \\
        \addlinespace
        Avg Gas/Tx & \textbf{-0.080***} & -0.001 & 13,373 & 0.170 \\
                 & (-7.26) & (-0.39) & & \\
        \midrule
        \multicolumn{5}{l}{\textbf{Panel B: AAI}} \\
        \addlinespace
        TVL & \textbf{-8.139***} & \textbf{2.252***} & 17,399 & 0.522 \\
            & (-37.00) & (8.79) & & \\
        \addlinespace
        Avg Gas/Tx & \textbf{1.036***} & \textbf{1.028***} & 13,373 & 0.163 \\
                 & (5.77) & (4.97) & & \\
        \midrule
        \multicolumn{5}{l}{Controls: Yes; Chain FE: Yes; Day FE: Yes} \\
        \bottomrule
      \end{tabular}
 
      \begin{tablenotes}[flushleft]
        \tiny 
        \setlength{\itemindent}{-2.5pt}
        \item \textit{Notes:} This table reports the heterogeneity of impact based on the bridge type.
        Columns (1)-(2) report coefficients for Offi (Official Bridges) and TP (Third-Party Bridges).
        $t$-statistics in parentheses.
        * p$<$0.1, ** p$<$0.05, *** p$<$0.01
      \end{tablenotes}
    \end{threeparttable}
  \end{minipage}
\hfill
  \begin{minipage}[t]{0.48\textwidth}
    \centering
    \begin{threeparttable}
      \caption{Heterogeneity Analysis by Cross-Chain Mechanism}
      \label{tab:heterogeneity_mechanism}
      \scriptsize
      \setlength{\tabcolsep}{2pt}
      \begin{tabular}{l c c c c c}
        \toprule
        Dep. Var & $A*I_{LNM}$ & $A*I_{BNM}$ & $A*I_{LP}$ & Obs & $R^2$ \\
        \midrule
        \multicolumn{6}{l}{\textbf{Panel A: ASI}} \\
        \addlinespace
        TVL & \textbf{0.044***} & \textbf{-0.033***} & \textbf{0.062***} & 17,399 & 0.496 \\
            & (15.14) & (-11.15) & (17.23) & & \\
        \addlinespace
        Avg Gas & \textbf{0.131***} & \textbf{-0.024***} & \textbf{-0.049***} & 13,373 & 0.239 \\
                 & (31.12) & (-8.24) & (-13.18) & & \\
        \midrule
        \multicolumn{6}{l}{\textbf{Panel B: AAI}} \\
        \addlinespace
        TVL & \textbf{0.888***} & \textbf{0.898***} & \textbf{0.938***} & 17,399 & 0.983 \\
            & (22.91) & (4.67) & (4.57) & & \\
        \addlinespace
        Avg Gas & \textbf{0.887***} & \textbf{1.595*} & \textbf{5.758***} & 13,373 & 0.164 \\
                 & (5.25) & (1.92) & (6.51) & & \\
        \midrule      
        \multicolumn{6}{l}{Controls: Yes; Chain FE: Yes; Day FE: Yes} \\
        \bottomrule
      \end{tabular}
      
      \begin{tablenotes}[flushleft]
        \tiny 
        \setlength{\itemindent}{-2.5pt}
        \item \textit{Notes:} This table reports estimated effects by bridging mechanism.
        Cols (1)-(3) are LNM (Lock-and-Mint), BNM (Burn-and-Mint), LP (Liquidity Pool).
        $t$-statistics in parentheses.
        * p$<$0.1, ** p$<$0.05, *** p$<$0.01
      \end{tablenotes}
    \end{threeparttable}
  \end{minipage}
\end{table}

\subsection{Design Implications for Cross-Chain Infrastructure}
Building on the heterogeneity results in Tables~\ref{tab:heterogeneity_bridge_type} and~\ref{tab:heterogeneity_mechanism}, we translate our measurement findings into three design implications.
A recurring theme is that deployed connectivity and realized flows represent distinct margins of interoperability, and they relate differently to execution costs and cross-chain coupling.

\begin{enumerate}
    \item \textbf{Design utilization-aware safeguards, not only broader coverage.}
    Across bridge types and mechanisms, higher active interoperability is associated with higher average gas cost per transaction, with the largest magnitude under liquidity-pool settlement in Table~\ref{tab:heterogeneity_mechanism}.
    This suggests that the externalities of interoperability are driven by flow surges rather than by connectivity alone.
    Utilization-scaled controls, such as governance-configurable rate limits and circuit breakers, can help bound tail risks during bursts and are already deployed in practice~\cite{osmosis_ibc_rate_limit,cosmos_circuit_docs}.
    Since many users route via aggregators, such controls can be implemented at the bridge layer or at routing and aggregation layers~\cite{metamask_bridge_aggregator}.

    \item \textbf{Differentiate canonical migration routes from external liquidity entry points.}
    Interoperability provided by official bridges is associated with lower gas cost per transaction and TVL patterns consistent with within-stack migration, while third-party bridge connectivity is associated with TVL increases consistent with external liquidity acquisition in Table~\ref{tab:heterogeneity_bridge_type}.
    These two roles serve different operational objectives.
    In practice, canonical routes can be optimized for moving assets between closely related chains, while third-party routes can be treated as entry points for external capital with separate routing and risk policies.

    \item \textbf{Avoid state-heavy settlement on corridors expected to carry frequent transfers.}
    Liquidity-pool designs exhibit the largest increase in gas-cost proxies when active flows rise in Table~\ref{tab:heterogeneity_mechanism}, consistent with higher per-transfer on-chain state updates under pool-based settlement.
    For corridors expected to carry frequent transfers, mechanisms with lower per-transfer state requirements may scale better, while liquidity-pool bridges are more suitable when routing flexibility is needed and volumes are moderate.
\end{enumerate}

% % % % % % % % % % % % % % % % % % % % % % % % % % % % % % % % % % % % % % % % % % % % 

\section{Related Work}
\label{sec:related}

Cross-chain interoperability has been studied from protocol, security, and empirical perspectives. Surveys systematize cross-chain communication and asset transfer mechanisms by verification assumptions, settlement models, and trust trade-offs~\cite{zamyatin2021sok,belchior2021survey,robinson2021survey,han2023survey,li2023review,li2025towards}, and a large body of work proposes new interoperability protocols, including atomic swaps, sidechains, payment-channel variants, and application-oriented frameworks~\cite{herlihy2018atomic,yin2021sidechains,tian2021enabling,jia2023cross,cao2025map,deng2025sfpow,sheng2023trustboost}. A complementary line of work studies bridge security, attack surfaces, architectural flaws, and failure modes in deployed systems~\cite{zhang2024security,augusto2024sok,notland2025sok,belenkov2025sok,zhang2022xscope,wu2025safeguarding}. Our paper approaches bridges from a different angle: the security literature asks how bridges fail and how failures can be mitigated, whereas we ask how bridge deployment and usage reshape ecosystem-level economic outcomes.

Recent empirical studies measure cross-chain transfers, costs, and routing infrastructures, often focusing on specific bridges, subsets of networks, or dataset construction for large-scale analysis~\cite{hu2024piecing,yan2025empirical,huang2024seamlessly,pillai2025wormhole,subramanian2024benchmarking,augusto2025xchaindatagen}. Among the studies most closely related to ours, Hu et al.~\cite{hu2024piecing} focus on reconstructing and profiling cross-chain transactions, including transaction purposes and abnormal activity. Augusto et al.~\cite{augusto2025xchaindatagen} primarily contribute a cross-chain dataset-generation framework and support protocol-level comparisons. Subramanian et al.~\cite{subramanian2024benchmarking} study routing infrastructures from the perspective of aggregator architecture, functionality, pricing, and latency. These studies provide valuable transaction-, protocol-, or routing-level evidence, but they do not separately measure the connectivity provisioned by deployed bridge infrastructure and the realized intensity of cross-chain usage. By contrast, we study bridges as economic infrastructure at the ecosystem level. We show that structural and active interoperability have distinct, sometimes opposite, associations with ecosystem growth, token returns, and congestion. These relationships further vary across bridge designs and chain architectures.

Our paper is also related to a growing literature on the economic consequences of DeFi infrastructure, including work on arbitrage, MEV, execution dynamics, and spillovers~\cite{oz2025cross,ferreira2024rolling,moosavi2023fast,kitzler2023disentangling,gramlich2023multivocal}. One line of work views blockchain ecosystems as linked networks and studies how infrastructure shapes ecosystem-level outcomes, such as oracle networks and inter-market connections~\cite{badev2023interconnected,cong2025financial} or layer-2 scaling~\cite{cong2023scaling}. Our view of bridges as cross-chain liquidity vehicles also connects to DeFi microstructure research on liquidity provision and execution~\cite{capponi2025liquidity,lehar2025decentralized,hasbrouck2025economic,hasbrouck2026need,capponi2026price}. We extend these largely single-chain analyses to cross-chain flows and show that capital-flow integration through bridges is associated with both ecosystem growth and higher fragility.

% % % % % % % % % % % % % % % % % % % % % % % % % % % % % % % % % % % % % % % % % % % % 
\section{Discussion and Conclusion}

\subsection{Summary of Findings}
This paper studies the economic consequences of cross-chain interoperability through bridge-mediated asset mobility. We construct a multi-source dataset covering 20 chains and 16 bridge protocols over 2022--2025 and model the ecosystem as a time-varying weighted hypergraph. This representation separates structural interoperability from active interoperability.
This separation turns out to matter. The corridor network densifies over time into a multi-hub core around EVM-compatible chains, but the consequences of this growing connectivity are not one-sided. Capital flows chase returns and avoid high costs, while infrastructure itself does not determine flow direction. Connectivity expands ecosystems but depresses native token returns at longer horizons, a pattern we call the Growth-Return Paradox. It also lowers average gas costs through load balancing, yet higher utilization adds congestion and tightens cross-chain co-movement, creating an Efficiency-Fragility Trade-off. These patterns are not uniform: they vary with chain architecture and bridge design, with official versus third-party bridges and different settlement mechanisms playing distinct roles.

\subsection{Novelty and Generalizability}

A common assumption in both industry and prior research is that more interoperability is better: wider bridge coverage means a more connected and efficient ecosystem. Our results complicate this picture. By separating what the infrastructure provides from how users actually use it, we show that the two dimensions sometimes pull in opposite directions. Infrastructure lowers gas costs but usage raises them, meaning the same bridge network acts as a load balancer or a congestion source depending on which dimension dominates. Infrastructure expands ecosystems in terms of TVL, users, and developer activity, but the resulting capital mobility depresses native token returns, and this effect is negligible at short horizons but grows steadily over weeks and months, pointing to a slow compositional shift rather than an immediate price response. Infrastructure connects chains, but that connection also synchronizes their economic cycles, and when a major bridge fails, exposed chains suffer disproportionate losses. These tensions are invisible when interoperability is measured as a single aggregate, which is why prior flow-based approaches have not detected them. The decomposition also reveals that bridges commonly grouped together in fact play different economic roles. Official bridges behave like vertical migration channels that move activity within a stack, while third-party bridges behave like horizontal entry points that attract outside liquidity. Liquidity pool settlement scales poorly under high usage, while lighter mechanisms handle volume better. These distinctions matter for practitioners choosing where to deploy and how to manage bridge infrastructure. Taken together, our findings suggest that interoperability is better understood as a design space with trade-offs than as a quantity to maximize.

A practical question is how far these results may travel beyond our sample. The measurement framework itself is portable. The hypergraph representation and the ASI/AAI decomposition work wherever on-chain bridge deployment data and transfer traces are available, which covers most public bridge ecosystems today. The boundary is off-chain activity. CEX-mediated cross-chain transfers, which involve internal netting on exchange ledgers, leave no corridor-level trace and fall outside our measurement scope.
The findings are also generalizable as Our sample is broad in both cross-section and time. It covers 20 chains spanning Layer-1, Layer-2, and sidechains, both EVM and non-EVM, connected by 16 bridges with diverse verification and settlement designs. Together these account for a large share of total bridge activity in the public blockchain ecosystem. The time window captures most bridges from near their inception and includes both bull and bear markets. This coverage, combined with identification strategies like the Multichain difference-in-differences, suggest that the qualitative patterns we document are robust. The specific magnitudes may naturally shift with market conditions and the arrival of new bridge designs. But the directional findings should not be sensitive to moderate changes in sample composition. These findings are patterns rooted in the structure of multi-chain systems, not in the particulars of any single market cycle.

\subsection{Implications for Cross-Chain Design and Security}

Our findings carry implications for both how bridges should be deployed and how their risks should be managed. Section \ref{sec:architecture} draws three design lessons from our heterogeneity results: manage utilization rather than only expanding coverage, differentiate canonical routes from third-party liquidity entry points, and match settlement mechanisms to expected corridor load. All three point away from treating interoperability as a uniform good and toward designing bridge deployments that are aware of what kind of traffic they will carry and how that traffic interacts with the local chain economy.

Our results also speak to bridge security, but from a different angle than the existing literature. Most security research focuses on direct attack vectors: validator compromise, contract exploits, and key management failures. These are important, but they are not the only way bridges create risk. Our contagion results show that structural connectivity synchronizes chain economies even when nothing is being exploited. Two chains linked by high-PSI bridges tend to move together, and when a major bridge fails, exposed chains lose TVL whether or not they were directly attacked. This is a form of systemic risk that protocol-level audits do not capture.
The two risk layers, direct exploits and indirect propagation, tend to concentrate on the same designs. Liquidity pool bridges are the most sensitive to utilization surges in our gas-cost regressions, and they are also the designs that pool user funds in ways that have historically attracted the largest exploits. This overlap has a practical implication. Congestion controls like rate limits and circuit breakers are usually discussed as efficiency tools. Our findings suggest they also serve as security infrastructure, because bounding flow surges during normal times limits the blast radius when things go wrong.

\subsection{Future Directions}
This paper provides a chain-day level analysis of how bridge infrastructure shapes blockchain economies. We believe two directions are worth exploring with finer-grained data. First, moving to wallet-level traces would allow studying individual bridger strategies, cross-chain liquidity provisions, and the micro mechanisms behind the aggregate patterns we document. Second, our findings raise a welfare question we do not answer: the Growth-Return Paradox implies that interoperability benefits some participants and hurts others, but who gains and who loses is not obvious. Ecosystem developers and new users may benefit from expanded activity, while native token holders bear the cost of diluted collateral composition. Making this precise would connect our measurement work to mechanism design questions about how bridge fees, incentives, and governance should be structured.

% % % % % % % % % % % % % % % % % % % % % % % % % % % % % % % % % % % % % % % % % % % % 
\begin{acks}
Siguang Li acknowledges the generous research support from the National Science Foundation of China (Grant No. 72573146). Xuechao Wang is supported by the Guangzhou-HKUST(GZ) Joint Funding Program (No. 2025A03J3882) and the
Guangzhou Municipal Science and Technology Project (No. 2025A04J4168). Lin William Cong is supported by NTU Global Institute of Finance, Technology, and Society (GIFTS).
\end{acks}

\bibliographystyle{ACM-Reference-Format}
\bibliography{ref}

@misc{yfinance,
  author = {V. Jose},
  title = {yfinance: Yahoo Finance market data downloader},
  year = {2020},
  howpublished = {\url{https://github.com/ranaroussi/yfinance}},
  note = {Accessed 2026-01-03}
}

@online{binance,
  author  = {Binance News},
  title   = {Cross-Chain Bridge Stargate's Volume Soars As Airdrop Hunters Set Sights on LayerZero Token},
  year    = {2023},
  howpublished     = {\url{https://www.binance.com/en-NG/square/post/467808}},
  note = {Accessed 2026-01-06}
}

@online{dlnews,
  author  = {Defillama News},
  title   = {Airdrop hunters send crypto bridge Stargate to new highs},
  year    = {2023},
  howpublished     = {\url{https://www.dlnews.com/articles/defi/airdrop-hunters-send-crypto-bridge-stargate-to-new-highs/}},
  note = {Accessed 2026-01-06}
}

@article{oz2025cross,
  title={Cross-chain arbitrage: The next frontier of mev in decentralized finance},
  author={{\"O}z, Burak and Torres, Christof Ferreira and Schlegel, Christoph and Mazorra, Bruno and Gebele, Jonas and Rezabek, Filip and Matthes, Florian},
  journal={Proceedings of the ACM on Measurement and Analysis of Computing Systems},
  volume={9},
  number={3},
  pages={1--33},
  year={2025},
  publisher={ACM New York, NY, USA}
}

@inproceedings{yan2025empirical,
  title={An Empirical Study on Cross-chain Transactions: Costs, Inconsistencies, and Activities},
  author={Yan, Kailun and Lu, Bo and Agrawal, Pranav and Li, Jiasun and Diao, Wenrui and Zhang, Xiaokuan},
  booktitle={Proceedings of the 20th ACM Asia Conference on Computer and Communications Security},
  pages={939--954},
  year={2025}
}

@article{augusto2025xchaindatagen,
  title={XChainDataGen: A Cross-Chain Dataset Generation Framework},
  author={Augusto, Andr{\'e} and Vasconcelos, Andr{\'e} and Correia, Miguel and Zhang, Luyao},
  journal={arXiv preprint arXiv:2503.13637},
  year={2025}
}

@inproceedings{zhang2024security,
  title={Security of cross-chain bridges: Attack surfaces, defenses, and open problems},
  author={Zhang, Mengya and Zhang, Xiaokuan and Zhang, Yinqian and Lin, Zhiqiang},
  booktitle={Proceedings of the 27th International Symposium on Research in Attacks, Intrusions and Defenses},
  pages={298--316},
  year={2024}
}

@article{badev2023interconnected,
  title={Interconnected DeFi: Ripple effects from the terra collapse},
  author={Badev, Anton I and Watsky, Cy},
  year={2023},
  publisher={FEDS Working Paper}
}

@techreport{cong2025financial,
  title={Financial and informational integration through oracle networks},
  author={Cong, Lin William and Prasad, Eswar S and Rabetti, Daniel},
  year={2025},
  institution={National Bureau of Economic Research}
}

@inproceedings{pillai2025wormhole,
  title={Wormhole Cross-Chain Bridge Transactions Flow: An Exploratory Study},
  author={Pillai, Babu and Tharani, Jeyakumar Samantha and Muthukkumarasamy, Vallipuram},
  booktitle={2025 IEEE International Conference on Blockchain and Cryptocurrency (ICBC)},
  pages={1--2},
  year={2025},
  organization={IEEE}
}

@inproceedings{huang2024seamlessly,
  title={Seamlessly Transferring Assets through Layer-0 Bridges: An Empirical Analysis of Stargate Bridge's Architecture and Dynamics},
  author={Huang, Chuanshan and Yan, Tao and Tessone, Claudio J},
  booktitle={Companion Proceedings of the ACM Web Conference 2024},
  pages={1776--1784},
  year={2024}
}

@article{hu2024piecing,
  title={Piecing Together the Jigsaw Puzzle of Transactions on Heterogeneous Blockchain Networks},
  author={Hu, Xiaohui and Feng, Hang and Xia, Pengcheng and Tyson, Gareth and Wu, Lei and Zhou, Yajin and Wang, Haoyu},
  journal={Proceedings of the ACM on Measurement and Analysis of Computing Systems},
  volume={8},
  number={3},
  pages={1--27},
  year={2024},
  publisher={ACM New York, NY, USA}
}

@article{belenkov2025sok,
  title={SoK: A review of cross-chain bridge hacks in 2023},
  author={Belenkov, Nikita and Callens, Valerian and Murashkin, Alexandr and Bak, Kacper and Derka, Martin and Gorzny, Jan and Lee, Sung-Shine},
  journal={arXiv preprint arXiv:2501.03423},
  year={2025}
}

@article{belchior2021survey,
  title={A survey on blockchain interoperability: Past, present, and future trends},
  author={Belchior, Rafael and Vasconcelos, Andr{\'e} and Guerreiro, S{\'e}rgio and Correia, Miguel},
  journal={Acm Computing Surveys (CSUR)},
  volume={54},
  number={8},
  pages={1--41},
  year={2021},
  publisher={ACM New York, NY}
}

@article{li2025towards,
  title={Towards Blockchain Interoperability: A Comprehensive Survey on Cross-Chain Solutions},
  author={Li, Wenqing and Liu, Zhenguang and Chen, Jianhai and Liu, Zhe and He, Qinming},
  journal={Blockchain: Research and Applications},
  pages={100286},
  year={2025},
  publisher={Elsevier}
}

@article{robinson2021survey,
  title={Survey of crosschain communications protocols},
  author={Robinson, Peter},
  journal={Computer Networks},
  volume={200},
  pages={108488},
  year={2021},
  publisher={Elsevier}
}

@inproceedings{zhang2022xscope,
  title={Xscope: Hunting for cross-chain bridge attacks},
  author={Zhang, Jiashuo and Gao, Jianbo and Li, Yue and Chen, Ziming and Guan, Zhi and Chen, Zhong},
  booktitle={Proceedings of the 37th IEEE/ACM International Conference on Automated Software Engineering},
  pages={1--4},
  year={2022}
}

@inproceedings{moosavi2023fast,
  title={Fast and furious withdrawals from optimistic rollups},
  author={Moosavi, Mahsa and Salehi, Mehdi and Goldman, Daniel and Clark, Jeremy},
  booktitle={5th Conference on Advances in Financial Technologies (AFT 2023)},
  pages={22--1},
  year={2023},
  organization={Schloss Dagstuhl--Leibniz-Zentrum f{\"u}r Informatik}
}

@article{li2023review,
  title={A review of blockchain cross-chain technology},
  author={Li, Li and Wu, Jiahao and Cui, Wei},
  journal={IET Blockchain},
  volume={3},
  number={3},
  pages={149--158},
  year={2023},
  publisher={Wiley Online Library}
}

@article{kitzler2023disentangling,
  title={Disentangling decentralized finance (DeFi) compositions},
  author={Kitzler, Stefan and Victor, Friedhelm and Saggese, Pietro and Haslhofer, Bernhard},
  journal={ACM Transactions on the Web},
  volume={17},
  number={2},
  pages={1--26},
  year={2023},
  publisher={ACM New York, NY}
}

@article{gramlich2023multivocal,
  title={A multivocal literature review of decentralized finance: Current knowledge and future research avenues},
  author={Gramlich, Vincent and Guggenberger, Tobias and Principato, Marc and Schellinger, Benjamin and Urbach, Nils},
  journal={Electronic Markets},
  volume={33},
  number={1},
  pages={11},
  year={2023},
  publisher={Springer}
}

@inproceedings{zamyatin2021sok,
  title     = {SoK: Communication across distributed ledgers},
  author    = {Zamyatin, Alexei and Al-Bassam, Mustafa and Zindros, Dionysis and Kokoris-Kogias, Eleftherios and Moreno-Sanchez, Pedro and Kiayias, Aggelos and Knottenbelt, William J.},
  booktitle = {Financial Cryptography and Data Security (FC)},
  series    = {LNCS},
  volume    = {12675},
  pages     = {3--36},
  publisher = {Springer},
  year      = {2021},
  doi       = {10.1007/978-3-662-64331-0_1}
}

@article{herlihy2018atomic,
  title   = {Atomic Cross-Chain Swaps},
  author  = {Herlihy, Maurice},
  journal = {arXiv preprint arXiv:1801.09515},
  year    = {2018}
}

@inproceedings{subramanian2024benchmarking,
  title={Benchmarking blockchain bridge aggregators},
  author={Subramanian, Shankar and Augusto, Andr{\'e} and Belchior, Rafael and Vasconcelos, Andr{\'e} and Correia, Miguel},
  booktitle={2024 IEEE International Conference on Blockchain (Blockchain)},
  pages={37--45},
  year={2024},
  organization={IEEE}
}

@article{han2023survey,
  title={A survey on cross-chain technologies},
  author={Han, Panpan and Yan, Zheng and Ding, Wenxiu and Fei, Shufan and Wan, Zhiguo},
  journal={Distributed ledger technologies: research and practice},
  volume={2},
  number={2},
  pages={1--30},
  year={2023},
  publisher={ACM New York, NY}
}

@article{notland2025sok,
  title={Sok: Cross-chain bridging architectural design flaws and mitigations},
  author={Notland, Jakob Svennevik and Li, Jingyue and Nowostawski, Mariusz and Haro, Peter Halland},
  journal={Blockchain: Research and Applications},
  pages={100315},
  year={2025},
  publisher={Elsevier}
}

@inproceedings{ferreira2024rolling,
  title={Rolling in the shadows: Analyzing the extraction of mev across layer-2 rollups},
  author={Ferreira Torres, Christof and Mamuti, Albin and Weintraub, Ben and Nita-Rotaru, Cristina and Shinde, Shweta},
  booktitle={Proceedings of the 2024 on ACM SIGSAC Conference on Computer and Communications Security},
  pages={2591--2605},
  year={2024}
}

@article{jia2023cross,
  title={Cross-chain virtual payment channels},
  author={Jia, Xiaofeng and Yu, Zhe and Shao, Jun and Lu, Rongxing and Wei, Guiyi and Liu, Zhenguang},
  journal={IEEE Transactions on Information Forensics and Security},
  volume={18},
  pages={3401--3413},
  year={2023},
  publisher={IEEE}
}

@article{tian2021enabling,
  title={Enabling cross-chain transactions: A decentralized cryptocurrency exchange protocol},
  author={Tian, Hangyu and Xue, Kaiping and Luo, Xinyi and Li, Shaohua and Xu, Jie and Liu, Jianqing and Zhao, Jun and Wei, David SL},
  journal={IEEE Transactions on Information Forensics and Security},
  volume={16},
  pages={3928--3941},
  year={2021},
  publisher={IEEE}
}

@inproceedings{cao2025map,
  title={Map the blockchain world: A trustless and scalable blockchain interoperability protocol for cross-chain applications},
  author={Cao, Yinfeng and Cao, Jiannong and Bai, Dongbin and Wen, Long and Liu, Yang and Li, Ruidong},
  booktitle={Proceedings of the ACM on Web Conference 2025},
  pages={717--726},
  year={2025}
}

@inproceedings{wu2025safeguarding,
  title={Safeguarding blockchain ecosystem: Understanding and detecting attack transactions on cross-chain bridges},
  author={Wu, Jiajing and Lin, Kaixin and Lin, Dan and Zhang, Bozhao and Wu, Zhiying and Su, Jianzhong},
  booktitle={Proceedings of the ACM on Web Conference 2025},
  pages={4902--4912},
  year={2025}
}

@article{deng2025sfpow,
  title={SFPoW: Constructing Secure and Flexible Proof-of-Work Sidechains for Cross-Chain Interoperability with Wrapped Assets},
  author={Deng, Zhihong and Tang, Chunming and Li, Taotao and Zeng, Zhikang and Abla, Parhat and He, Debiao},
  journal={IEEE Transactions on Computers},
  year={2025},
  publisher={IEEE}
}

@article{yin2021sidechains,
  title={Sidechains with fast cross-chain transfers},
  author={Yin, Lingyuan and Xu, Jing and Tang, Qiang},
  journal={IEEE Transactions on Dependable and Secure Computing},
  volume={19},
  number={6},
  pages={3925--3940},
  year={2021},
  publisher={IEEE}
}

@inproceedings{augusto2024sok,
  title={Sok: Security and privacy of blockchain interoperability},
  author={Augusto, Andr{\'e} and Belchior, Rafael and Correia, Miguel and Vasconcelos, Andr{\'e} and Zhang, Luyao and Hardjono, Thomas},
  booktitle={2024 IEEE Symposium on Security and Privacy (SP)},
  pages={3840--3865},
  year={2024},
  organization={IEEE}
}

@inproceedings{sheng2023trustboost,
  title={Trustboost: Boosting trust among interoperable blockchains},
  author={Sheng, Peiyao and Wang, Xuechao and Kannan, Sreeram and Nayak, Kartik and Viswanath, Pramod},
  booktitle={Proceedings of the 2023 ACM SIGSAC Conference on Computer and Communications Security},
  pages={1571--1584},
  year={2023}
}

@misc{eth_optimistic_rollups,
  author       = {{Ethereum Foundation}},
  title        = {Optimistic rollups},
  year         = {2024},
  howpublished = {\url{https://ethereum.org/developers/docs/scaling/optimistic-rollups/}},
  note         = {Accessed 2026-01-13}
}

@misc{eth_zk_rollups,
  author       = {{Ethereum Foundation}},
  title        = {Zero-knowledge rollups},
  year         = {2024},
  howpublished = {\url{https://ethereum.org/developers/docs/scaling/zk-rollups/}},
  note         = {Accessed 2026-01-13}
}

@misc{dune,
  author       = {{Dune Analytics}},
  title        = {Dune Analytics},
  year         = {2026},
  howpublished = {\url{https://dune.com/home}},
  note         = {Accessed 2026-01-13}
}

@misc{defillama_chain_tvl_api,
  author       = {{DefiLlama}},
  title        = {DefiLlama API: Historical Chain TVL},
  year         = {2026},
  howpublished = {\url{https://api.llama.fi/v2/historicalChainTvl/}},
  note         = {Accessed 2026-01-13}
}

@misc{eip4844,
  author       = {{Ethereum Improvement Proposals}},
  title        = {EIP-4844: Shard Blob Transactions},
  year         = {2022},
  howpublished          = {\url{https://eips.ethereum.org/EIPS/eip-4844}},
  note         = {Accessed 2026-01-10}
}

@misc{dln_stats_api,
  author       = {{deBridge}},
  year         = {2026},
  title        = {deBridge DLN API: Daily Statistics Endpoint},
  howpublished = {\url{https://dln-api.debridge.finance/api/Satistics/getDaily}},
  note         = {Accessed 2026-01-13}
}

@misc{dln_orders_api,
  author       = {{deBridge}},
  year         = {2026},
  title        = {deBridge DLN API: Orders Filtered List Endpoint},
  howpublished = {\url{https://stats-api.dln.trade/api/Orders/filteredList}},
  note         = {Accessed 2026-01-13}
}

@misc{defillama_protocol_api,
  author       = {{DefiLlama}},
  year         = {2026},
  title        = {DefiLlama API: Protocol TVL},
  howpublished = {\url{https://api.llama.fi/protocol}},
  note         = {Accessed 2026-01-13}
}

@misc{wormholescan_api,
  author       = {{Wormhole Foundation}},
  year         = {2026},
  title        = {WormholeScan API: Operations Endpoint},
  howpublished = {\url{https://api.wormholescan.io/api/v1/operations}},
  note         = {Accessed 2026-01-13}
}

@misc{osmosis_ibc_rate_limit,
  author       = {{Osmosis Labs}},
  year         = {2026},
  title        = {IBC Rate Limit Module},
  howpublished = {\url{https://github.com/osmosis-labs/osmosis/blob/main/x/ibc-rate-limit/README.md}},
  note         = {Accessed 2026-01-13}
}

@misc{cosmos_circuit_docs,
  author       = {{Cosmos SDK}},
  year         = {2026},
  title        = {x/circuit: Circuit Breaker Module Documentation},
  howpublished = {\url{https://docs.cosmos.network/sdk/v0.50/build/modules/circuit/README}},
  note         = {Accessed 2026-01-13}
}

@misc{metamask_bridge_aggregator,
  author       = {{MetaMask}},
  year         = {2026},
  title        = {MetaMask Launches Bridge Aggregator in dApp},
  howpublished = {\url{https://metamask.io/news/metamask-launches-bridge-aggregator-in-dapp-to-easily-move-tokens-across-chains}},
  note         = {Accessed 2026-01-13}
}

@misc{defillama_bridges,
  author       = {{DefiLlama}},
  year         = {2026},
  title        = {DefiLlama Bridges Dashboard},
  howpublished = {\url{https://defillama.com/bridges}},
  note         = {Accessed 2026-01-13}
}

@misc{wormholescan_upgrade_blog,
  author       = {{Wormhole}},
  year         = {2026},
  title        = {WormholeScan Upgrade: Real-Time Data Analytics for the Wormhole Ecosystem},
  howpublished = {\url{https://wormhole.com/blog/wormholescan-upgrade-real-time-data-analytics-for-the-wormhole-ecosystem}},
  note         = {Accessed 2026-01-13}
}

@article{cong2023scaling,
  title={Scaling smart contracts via layer-2 technologies: Some empirical evidence},
  author={Cong, Lin William and Hui, Xiang and Tucker, Catherine and Zhou, Luofeng},
  journal={Management Science},
  volume={69},
  number={12},
  pages={7306--7316},
  year={2023},
  publisher={INFORMS}
}

@article{capponi2025liquidity,
  title={Liquidity provision on blockchain-based decentralized exchanges},
  author={Capponi, Agostino and Jia, Ruizhe},
  journal={The Review of Financial Studies},
  volume={38},
  number={10},
  pages={3040--3085},
  year={2025},
  publisher={Oxford University Press}
}

@article{lehar2025decentralized,
  title={Decentralized exchange: The uniswap automated market maker},
  author={Lehar, Alfred and Parlour, Christine},
  journal={The Journal of Finance},
  volume={80},
  number={1},
  pages={321--374},
  year={2025},
  publisher={Wiley Online Library}
}

@article{hasbrouck2025economic,
  title={An economic model of a decentralized exchange with concentrated liquidity},
  author={Hasbrouck, Joel and Rivera, Thomas J and Saleh, Fahad},
  journal={Management Science},
  year={2025},
  publisher={INFORMS}
}

@article{hasbrouck2026need,
  title={The need for fees at a dex: How increases in fees can increase dex trading volume},
  author={Hasbrouck, Joel and Rivera, Thomas J and Saleh, Fahad},
  journal={Management Science},
  year={2026},
  publisher={INFORMS}
}

@article{capponi2026price,
  title={Price discovery on decentralized exchanges},
  author={Capponi, Agostino and Jia, Ruizhe and Yu, Shihao},
  journal={The Review of Financial Studies},
  pages={hhag002},
  year={2026},
  publisher={Oxford University Press}
}

\appendix

\section{Dune-Based Extraction Details}
\label{app:dune}

Dune is used as the primary on-chain measurement back-end for (i) chain-level daily attributes and (ii) bridge-level daily flows when the corresponding contracts and event tables are indexed and queryable. Dune continuously ingests raw blockchain data (blocks, transactions, logs, traces) for many networks and exposes them through a unified SQL interface, while also maintaining curated, structured datasets such as decoded event tables (e.g., \texttt{erc20\_<chain>.evt\_transfer}) and standardized price feeds (e.g., \texttt{prices.minute}). All Dune queries in this study follow a common template: we define an explicit time window via a \texttt{time\_filter} CTE; we normalize timestamps using \texttt{DATE\_TRUNC} to the target granularity; we filter to successful transactions or valid transfers; and we aggregate to daily panels with consistent field names. Representative examples include (a) gas cost queries that compute per-transaction gas expenditure as \texttt{gas\_used} $\times$ \texttt{gas\_price} $\times 10^{-18}$(here number 18 can be replaced by other corresponding decimals) and join \texttt{prices.minute} at minute resolution to obtain USD-denominated daily totals and distributional statistics, (b) daily active user queries that count distinct transaction senders (\texttt{COUNT(DISTINCT from)}) among successful transactions, and (c) contract deployment queries that count distinct created contract addresses from creation trace tables where available. 

For bridge-level daily flows on Dune, we identify bridge-mediated transfers using decoded event tables and/or ERC20 transfer logs. When a bridge exposes a clear escrow/router address on a given chain, we extract inbound deposits and outbound withdrawals by filtering \texttt{erc20\_<chain>.evt\_transfer} events where the bridge address appears as \texttt{to} (inbound) or \texttt{from} (outbound), and we restrict to a corridor-specific token list (e.g., stablecoins or canonical assets used by that bridge for the corridor). We then aggregate to daily (bridge, source chain, destination chain) statistics, including \texttt{transfer\_count} (\texttt{COUNT(*)}), \texttt{daily\_users} (\texttt{COUNT(DISTINCT from)} or \texttt{COUNT(DISTINCT to)} under a consistent direction convention), and transfer-size summaries such as mean and quantiles computed via \texttt{AVG} and \texttt{APPROX\_PERCENTILE}. The opBNB canonical bridge query illustrates this logic by extracting deposits to the bridge escrow address and withdrawals from it, then combining both directions via \texttt{UNION ALL} to form a corridor-day panel. When USD conversion is required and a reliable on-chain price feed is available, we join \texttt{prices.minute} at minute resolution and aggregate to daily USD notional; otherwise, for USD-pegged stablecoin corridors we treat token units (after decimals) as a USD proxy. All Dune outputs are exported as CSV and post-processed to normalize chain identifiers, align to UTC day boundaries, and match the unified schema used for integrating API-derived flows (Wormhole and deBridge).

\section{Covered Chains and Bridges}
Tables~\ref{tab:chains_covered} and~\ref{tab:bridges_covered} summarize the chains and bridge entities covered in our dataset, including each chain's execution stack and ecosystem, the availability of a canonical routing option, and high-level bridge design attributes such as settlement model and verification assumptions.

\begin{table}[t]
\centering
\footnotesize
\setlength{\tabcolsep}{4pt}
\renewcommand{\arraystretch}{1.15}
\begin{tabularx}{\linewidth}{
    >{\raggedright\arraybackslash}p{0.12\linewidth}
    >{\centering\arraybackslash}p{0.10\linewidth}
    >{\centering\arraybackslash}p{0.2\linewidth}
    >{\centering\arraybackslash}p{0.1\linewidth}
    >{\raggedleft\arraybackslash}p{0.1\linewidth}
    >{\raggedright\arraybackslash}X
}
\toprule
Chain & Stack & VM / Ecosystem & Canonical bridge & Median TVL & Notes \\
\midrule
Aptos  & Layer 1 & Move & No & \$355.97M & Move-based L1 \\
Avalanche & Layer 1 & EVM & Yes & \$1.05B & Canonical bridge to Ethereum \\
BNB Chain & Layer 1 & EVM & Yes & \$5.42B & BNB ecosystem; canonical bridge to opBNB/BNB Beacon \\
Ethereum & Layer 1 & EVM & No & \$26.87B & Major settlement hub \\
Sei& Layer 1 & Cosmos (IBC) & IBC & \$169.72M & IBC-enabled corridor(s) \\
Solana & Layer 1 & SVM & No & \$3.77B & Solana VM; often via third-party bridges \\
Sonic & Layer 1 & EVM & Yes & \$341.97M & EVM L1 \\
Sui & Layer 1 & Move & No & \$678.34M & Move-based L1 \\
TRON & Layer 1 & TVM & Yes & \$5.32B & TRON VM \\
\addlinespace
Arbitrum One & Layer 2 & EVM (Optimistic rollup) & Yes & \$2.21B & Canonical bridge to Ethereum \\
Base & Layer 2 & EVM (OP Stack) & Yes & \$1.64B & Canonical bridge to Ethereum \\
Linea & Layer 2 & EVM (ZK rollup) & Yes & \$216.89M & Canonical bridge to Ethereum \\
opBNB  & Layer 2 & EVM (OP Stack) & Yes & \$18.56M & Canonical bridge to BNB Chain \\
Optimism & Layer 2 & EVM (OP Stack) & Yes & \$621.09M & Canonical bridge to Ethereum \\
Unichain  & Layer 2 & EVM & No & \$262.01M & L2 (canonical route not specified) \\
Hyperliquid & Layer 1 & Custom (HyperEVM) & Yes & -- & Native bridge to Arbitrum corridor \\
\addlinespace
Cronos & Sidechain & Cosmos + EVM & Yes & \$447.90M & EVM-compatible Cosmos chain \\
Polygon PoS & Sidechain & EVM & Yes & \$1.02B & PoS sidechain; canonical bridge to Ethereum \\
Ronin & Sidechain & EVM & Yes & \$84.54M & Gaming sidechain; canonical bridge to Ethereum \\
Rootstock & Sidechain & BTC + EVM & Yes & \$127.95M & Bitcoin-linked EVM sidechain \\
\bottomrule
\end{tabularx}
\caption{Chains covered in this study.
``Canonical bridge'' indicates whether a widely used canonical/native route exists for the chain (IBC for Cosmos-IBC chains).
Median TVL is computed from \texttt{chain\_attributes\_addRonin.csv} when available; `--' indicates missing/unavailable TVL in the chain-attribute dataset.}
\label{tab:chains_covered}
\end{table}

\begin{table}[t]
\centering
\footnotesize
\setlength{\tabcolsep}{3pt}
\renewcommand{\arraystretch}{1.15}
\begin{tabularx}{\linewidth}{l l >{\raggedright\arraybackslash}X >{\raggedright\arraybackslash}X c}
\toprule
Bridge (dataset id) & Category & Asset handling / model & Verification (high-level) & Created \\
\midrule
across & Third-party & Liquidity pool / intent-style routing & Optimistic verification (UMA OO) + relayers & 2021-11 \\
arbitrum bridge & Canonical & Lock-and-mint & Fraud proofs (optimistic rollup) & 2021-08 \\
base standard bridge & Canonical & Lock-and-mint & Fraud proofs (OP Stack) & 2023-08 \\
cctp & Token protocol & Burn-and-mint (USDC) & Circle attestation / issuer-controlled & 2023-04 \\
~cctp v1 & Token protocol & Burn-and-mint (USDC) & Circle attestation / issuer-controlled & 2023-04 \\
~cctp v2 & Token protocol & Burn-and-mint (USDC) & Circle attestation / issuer-controlled & 2025-03 \\
debridge & Third-party & Lock-and-mint (DLN) / liquidity network & PoS validator set / gateway & 2023-06 \\
hyperliquid bridge & Canonical & Lock-and-mint & PoA committee / multisig & 2023-12 \\
layerzero & Underlying protocol & Depends on apps (lock/mint or pool-based) & DVN (oracle + relayer) & 2022-03 \\
~Stargate & Underlying protocol & Depends on apps (lock/mint or pool-based) & DVN (oracle + relayer) & 2022-03 \\
~USDT0 & Underlying protocol & Depends on apps (lock/mint or pool-based) & DVN (oracle + relayer) & 2025-01 \\
opbnb native bridge & Canonical & Lock-and-mint & OP Stack / rollup-based & 2023-09 \\
optimism bridge & Canonical & Lock-and-mint & Fraud proofs (optimistic rollup) & 2021-12 \\
polygon bridge & Canonical & Lock-and-mint & PoS validators + checkpoint & 2020-05 \\
ronin & Canonical & Lock-and-mint & Validator committee / multisig (Ronin) & 2021-02 \\
wormhole & Third-party & Lock-and-mint (message + token transfer) & Guardian set (PoA) & 2021-08 \\
\bottomrule
\end{tabularx}
\caption{Bridges covered in the bridge-flow dataset. Created dates indicate first public availability (some are month-level due to disclosure granularity).}
\label{tab:bridges_covered}
\end{table}

\section{Across-Window Distributions from Daily Quantile Summaries}
\label{app:quantile_mixing}

Our core panels are daily aggregates rather than per-transaction traces.
To report latency, fee, and value distributions in a format comparable to transaction-level studies, we approximate the per-transfer distribution over a reporting window by aggregating day-level quantile summaries.
For each bridge--corridor on day $t$, we record a five-number summary for a metric, consisting of the minimum, first quartile, median, third quartile, and maximum.
We interpret these quantiles as defining a piecewise-linear cumulative distribution function for that day.
We then form a window-level mixture by weighting each day's CDF by its transfer count, so days with more transfers contribute proportionally more mass.
Window-level quantiles and the interquartile range are obtained by inverting the resulting count-weighted mixture CDF.
This procedure uses the full set of daily quantile fields while avoiding storage of individual transactions.

A fully disaggregated bridge by corridor distribution table is prohibitively large at our scale.
Over the full window from 2022--01--01 to 2025--10--31, we observe 378 directed corridors and 1{,}051 bridge--corridor pairs with nonzero activity.
Reporting corridor-level distributions would therefore span many pages and hinder readability.
Instead, we provide compact bridge-level and chain-level distributional summaries that preserve comparability and reflect the availability of latency, fee, and value attributes across entities.
The resulting bridge-level and chain-level summaries are reported in Tables~\ref{tab:app_bridge_compact_multi} and~\ref{tab:app_chain_compact_multi}.

\begin{table*}[t]
\centering
\scriptsize
\setlength{\tabcolsep}{2pt}
\renewcommand{\arraystretch}{1.08}
\caption{Compact bridge-level distribution summaries over the full study window (2022-01-01 to 2025-10-31). Each metric reports $n$ total transfers and $d$ active days as $n/d$, followed by Q1/Q2/Q3 and IQR. Quantiles are computed from a transfer-count-weighted mixture of day-level five-number summaries (min/Q1/median/Q3/max) using a piecewise-linear CDF approximation.}
\label{tab:app_bridge_compact_multi}
\begin{tabular}{lccccccccccccccc}
\toprule
\multirow{2}{*}{Bridge} & \multicolumn{5}{c}{Transfer value (USD)} & \multicolumn{5}{c}{Fee (USD)} & \multicolumn{5}{c}{Latency (s)} \\
\cmidrule(lr){2-6}\cmidrule(lr){7-11}\cmidrule(lr){12-16}
 & $n/d$ & Q1 & Q2 & Q3 & IQR & $n/d$ & Q1 & Q2 & Q3 & IQR & $n/d$ & Q1 & Q2 & Q3 & IQR \\
\midrule
layerzero & 66.0M/1325 & 5.15 & 60.22 & 1,054 & 1,049 & 66.0M/1325 & 0.343 & 0.606 & 1.761 & 1.418 & 33.4M/823 & 33 & 55 & 137 & 104 \\
stargate & 49.3M/1325 & 10.90 & 82.65 & 1,095 & 1,084 & 49.3M/1325 & 0.343 & 0.607 & 1.487 & 1.145 & 25.0M/804 & 32 & 51 & 110 & 77 \\
usdt0 & 237.3K/276 & 64.97 & 865.86 & 22,161 & 22,096 & 237.3K/276 & 0.022 & 0.065 & 0.667 & 0.645 & 237.3K/276 & 49 & 107 & 480 & 431 \\
across & 13.9M/1400 & 10.84 & 64.01 & 1,423 & 1,412 & 13.8M/1400 & 0.013 & 0.076 & 1.836 & 1.823 & -- & -- & -- & -- & -- \\
hyperliquid native bridge & 4.8M/699 & 90.64 & 542.8 & 10,881 & 10,791 & -- & -- & -- & -- & -- & -- & -- & -- & -- & -- \\
wormhole & 3.5M/1400 & 20.63 & 235.9 & 1,987 & 1,967 & 1.3M/425 & 0.001 & 0.006 & 0.543 & 0.541 & 3.3M/1041 & 93 & 1,030 & 1,453 & 1,360 \\
polygon native bridge & 1.3M/1400 & 31.49 & 209.5 & 26,817 & 26,786 & -- & -- & -- & -- & -- & -- & -- & -- & -- & -- \\
base standard bridge & 1.3M/851 & 22.00 & 140.9 & 6,372 & 6,350 & -- & -- & -- & -- & -- & -- & -- & -- & -- & -- \\
arbitrum native bridge & 1.1M/1400 & 18.15 & 101.3 & 6,983 & 6,965 & -- & -- & -- & -- & -- & -- & -- & -- & -- & -- \\
cctp\_both & 1.0M/304 & 46.10 & 1,234 & 31,146 & 31,099 & -- & -- & -- & -- & -- & -- & -- & -- & -- & -- \\
cctp\_v1 & 711.1K/304 & 31.65 & 934.3 & 34,566 & 34,534 & -- & -- & -- & -- & -- & -- & -- & -- & -- & -- \\
% base native bridge & 446.6K/846 & 18.13 & 116.6 & 2,825 & 2,807 & -- & -- & -- & -- & -- & -- & -- & -- & -- & -- \\
optimism native bridge & 419.3K/1400 & 14.33 & 84.43 & 34,384 & 34,369 & -- & -- & -- & -- & -- & -- & -- & -- & -- & -- \\
cctp\_v2 & 320.7K/244 & 82.32 & 2,312 & 21,107 & 21,025 & -- & -- & -- & -- & -- & -- & -- & -- & -- & -- \\
opbnb native bridge & 30.1K/779 & 13.87 & 75.68 & 556.6 & 542.7 & -- & -- & -- & -- & -- & -- & -- & -- & -- & -- \\
debridge & 5.6M/1028 & 425.4 & 1,168 & 3,342 & 2,916 & 5.6M/1028 & 3.049 & 3.586 & 4.509 & 1.460 & -- & -- & -- & -- & -- \\
ronin & 320.1K/1200 & 4,080 & 9,481 & 24,801 & 20,721 & -- & -- & -- & -- & -- & -- & -- & -- & -- & -- \\
\bottomrule
\end{tabular}
\end{table*}

\begin{table*}[t]
\centering
\scriptsize
\setlength{\tabcolsep}{3pt}
\renewcommand{\arraystretch}{1.08}
\caption{Compact chain-level endpoint distribution summaries over the full study window (2022-01-01 to 2025-10-31). Each metric reports $n$ total endpoint-attributed transfers and $d$ active days as $n/d$, followed by Q1/Q2/Q3 and IQR. Quantiles are computed from a transfer-count-weighted mixture of day-level five-number summaries (min/Q1/median/Q3/max) using a piecewise-linear CDF approximation.}
\label{tab:app_chain_compact_multi}
\begin{tabular}{lccccccccccccccc}
\toprule
\multirow{2}{*}{Chain} & \multicolumn{5}{c}{Transfer value (USD)} & \multicolumn{5}{c}{Fee (USD)} & \multicolumn{5}{c}{Latency (s)} \\
\cmidrule(lr){2-6}\cmidrule(lr){7-11}\cmidrule(lr){12-16}
 & $n/d$ & Q1 & Q2 & Q3 & IQR & $n/d$ & Q1 & Q2 & Q3 & IQR & $n/d$ & Q1 & Q2 & Q3 & IQR \\
\midrule
arbitrum & 44.9M/1400 & 10.55 & 92.86 & 2,478 & 2,468 & 37.8M/1400 & 0.252 & 0.697 & 2.141 & 1.889 & 16.4M/925 & 27 & 41 & 92 & 64 \\
optimism & 26.5M/1400 & 8.31 & 71.11 & 1,035 & 1,027 & 25.5M/1400 & 0.083 & 0.700 & 1.739 & 1.656 & 11.6M/884 & 37 & 53 & 87 & 50 \\
base & 23.3M/853 & 9.41 & 71.19 & 1,541 & 1,531 & 20.3M/837 & 0.051 & 0.180 & 1.494 & 1.443 & 12.9M/824 & 31 & 47 & 115 & 84 \\
polygon & 22.7M/1400 & 3.43 & 45.48 & 861.9 & 858.5 & 20.6M/1325 & 0.341 & 0.420 & 1.090 & 0.749 & 6.6M/967 & 64 & 531 & 1,202 & 1,138 \\
bsc & 17.7M/1399 & 4.30 & 59.07 & 1,030 & 1,025 & 17.2M/1325 & 0.418 & 0.619 & 1.382 & 0.964 & 7.3M/964 & 45 & 76 & 147 & 102 \\
avalanche & 17.5M/1374 & 4.70 & 62.01 & 1,405 & 1,401 & 16.9M/1325 & 0.410 & 0.512 & 1.317 & 0.907 & 5.2M/965 & 37 & 51 & 113 & 76 \\
ethereum & 13.0M/1400 & 31.00 & 224.0 & 11,718 & 11,687 & 7.5M/1400 & 0.086 & 1.144 & 8.894 & 8.807 & 2.6M/1009 & 101 & 212 & 893 & 792 \\
aptos & 7.2M/1109 & 0.16 & 5.07 & 883.9 & 883.8 & 7.0M/1109 & 0.272 & 0.440 & 1.070 & 0.798 & 3.1M/965 & 58 & 82 & 316 & 259 \\
linea & 6.6M/837 & 15.54 & 66.89 & 1,109 & 1,094 & 6.6M/837 & 0.074 & 0.298 & 1.986 & 1.913 & 4.5M/801 & 37 & 61 & 146 & 108 \\
hyperliquid & 5.0M/699 & 87.39 & 529.6 & 8,859 & 8,772 & 162.9K/190 & 0.022 & 0.053 & 0.155 & 0.133 & 143.6K/190 & 43 & 60 & 717 & 674 \\
solana & 1.7M/1400 & 85.95 & 409.6 & 2,862 & 2,776 & 707.3K/472 & 0.079 & 0.526 & 2.840 & 2.761 & 1.6M/814 & 80 & 903 & 1,377 & 1,296 \\
unichain & 1.1M/268 & 4.76 & 44.51 & 3,868 & 3,863 & 1.0M/267 & 0.004 & 0.021 & 0.108 & 0.104 & 346.6K/263 & 35 & 46 & 79 & 44 \\
opbnb & 365.9K/782 & 0.01 & 0.14 & 3.48 & 3.47 & 335.8K/776 & 0.115 & 0.195 & 1.342 & 1.227 & 335.8K/776 & 36 & 76 & 1,197 & 1,161 \\
sonic & 352.0K/318 & 401.9 & 2,119 & 11,184 & 10,782 & 196.8K/318 & 0.089 & 0.115 & 0.481 & 0.392 & 196.8K/318 & 30 & 47 & 67 & 37 \\
sui & 330.8K/914 & 65.22 & 426.1 & 4,565 & 4,500 & 28.2K/425 & 0.002 & 0.181 & 2.585 & 2.583 & 271.2K/906 & 23 & 83 & 1,100 & 1,077 \\
worldchain & 316.3K/387 & 5.20 & 46.16 & 630.1 & 624.9 & 224.7K/387 & 0.007 & 0.055 & 0.351 & 0.345 & 3.4K/209 & 74 & 77 & 80 & 6 \\
sei & 220.1K/822 & 11.63 & 172.5 & 3,577 & 3,565 & 160.1K/504 & 0.021 & 0.049 & 0.072 & 0.051 & 166.1K/672 & 18 & 32 & 65 & 48 \\
tron & 21.1K/296 & 700.8 & 6,443 & 60,521 & 59,820 & 21.1K/296 & 6.794 & 25.347 & 27.775 & 20.981 & 21.1K/296 & 50 & 90 & 215 & 165 \\
rootstock & 7.4K/302 & 52.37 & 1,597 & 12,849 & 12,797 & 7.4K/302 & 0.541 & 1.260 & 1.466 & 0.925 & 7.4K/302 & 315 & 501 & 604 & 290 \\
ronin & 320.1K/1200 & 4,080 & 9,481 & 24,801 & 20,721 & -- & -- & -- & -- & -- & -- & -- & -- & -- & -- \\
\bottomrule
\end{tabular}
\end{table*}

\section{Bridge-Level Count--Notional Asymmetry}
\label{app:bridge_count_amount_asym}

This appendix provides complementary views of the count--notional asymmetry discussed in Section~4.2.
We summarize how bridges differ in (i) their aggregate shares over the full study window and (ii) how the imbalance evolves over time.
Throughout, a bridge's \emph{count share} is its fraction of total transfer counts, and its \emph{amount share} is its fraction of total bridged USD notional, computed over the same population of transfers.

\begin{figure}[t]
    \centering
    \includegraphics[width=0.5\linewidth]{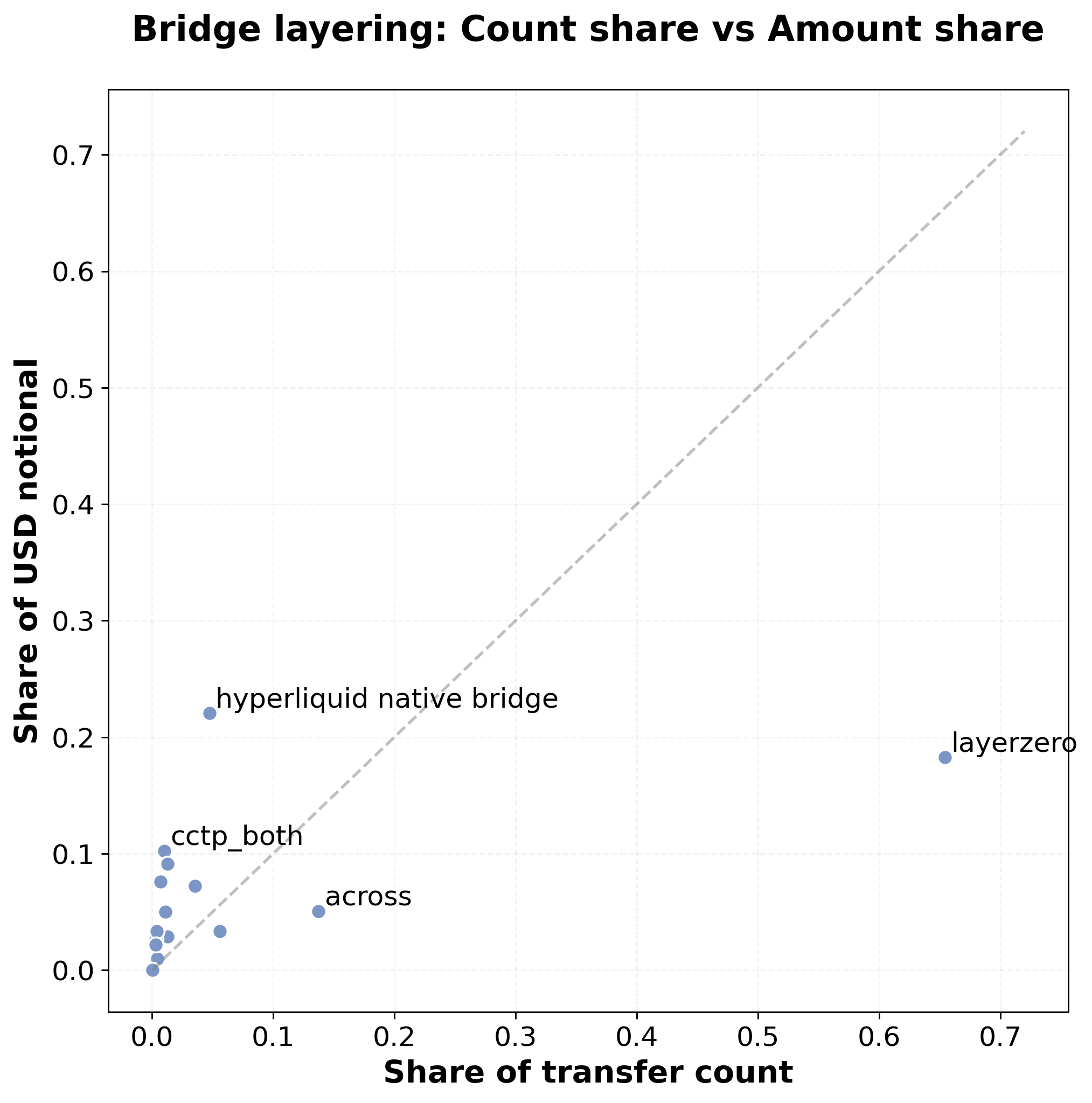}
    \caption{Bridge ``layering'' by overall share of transfer counts (x-axis) versus overall share of USD notional (y-axis) over the study window. The diagonal $y=x$ indicates symmetry; points above (below) the diagonal are value-heavy (count-heavy).}
    \label{fig:bridge_layering_scatter}
\end{figure}

\paragraph{Cross-sectional ``layering'' across bridges.}
Figure~\ref{fig:bridge_layering_scatter} plots each bridge by its overall count share (x-axis) and overall amount share (y-axis).
The diagonal $y=x$ marks symmetry: points above the diagonal are \emph{value-heavy} (higher amount share than count share), whereas points below are \emph{count-heavy} (higher count share than amount share).
Consistent with Section~4.2, bridges such as the Hyperliquid native bridge appear value-heavy, while LayerZero-based routes appear count-heavy.

\begin{figure}[t]
    \centering
    \includegraphics[width=0.99\linewidth]{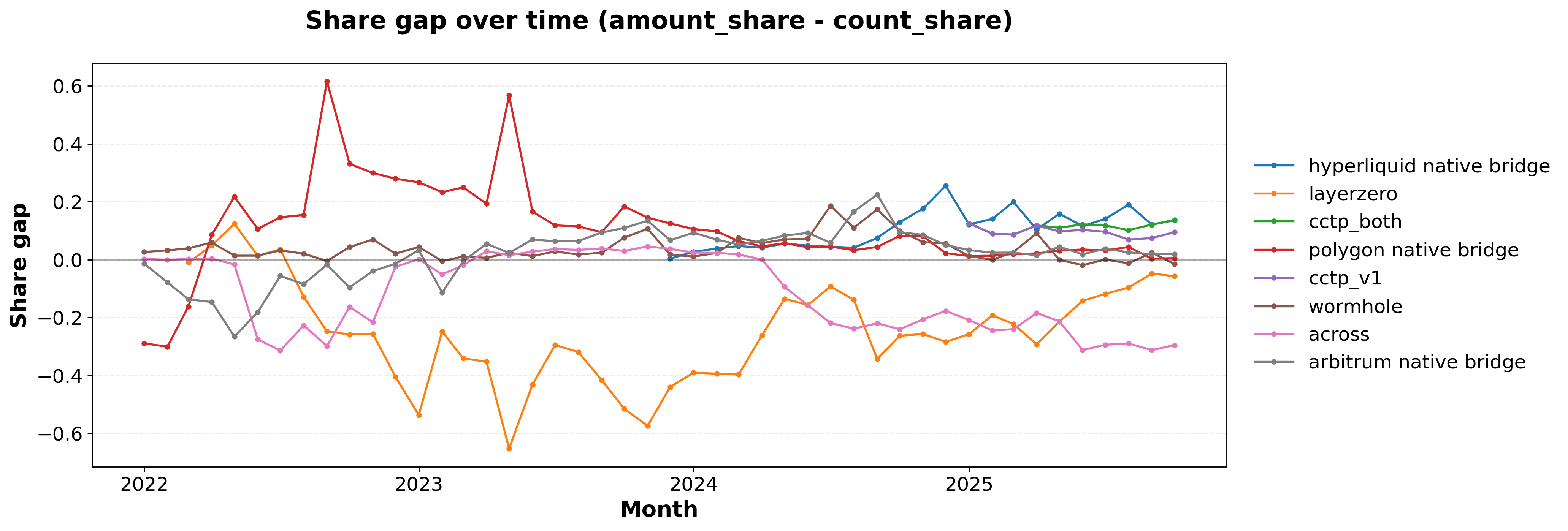}
    \caption{Monthly share gap $\Delta_b(t)=s^{\text{amount}}_b(t)-s^{\text{count}}_b(t)$ for the top-8 bridges by aggregate USD notional. Positive values indicate that a bridge accounts for a larger share of notional than of counts (value-heavy), and negative values indicate the opposite (count-heavy).}
    \label{fig:bridge_share_gap_time}
\end{figure}

\paragraph{Dynamics of the share gap over time.}
To capture temporal variation, we define the \emph{share gap} for bridge $b$ at month $t$ as
$\Delta_b(t)= s^{\text{amount}}_b(t)-s^{\text{count}}_b(t)$, where $s^{\text{amount}}_b(t)$ and $s^{\text{count}}_b(t)$ are the monthly shares of USD notional and transfer counts, respectively.
Figure~\ref{fig:bridge_share_gap_time} plots $\Delta_b(t)$ for the top-8 bridges ranked by aggregate USD notional.
A positive (negative) gap indicates that a bridge carries disproportionately large (small) transfers relative to its usage intensity.
The figure highlights that the direction and magnitude of imbalance can be persistent for some bridges yet time-varying for others, complementing the weekly composition plots in Figure~\ref{fig:weekly_bridge_activity}.

\clearpage
\section{Omitted Regression Tables}

Table~\ref{tab:pearson_corr} provides unconditional correlations between the interoperability metrics and economic indicators. Most correlations are statistically significant, but the magnitudes vary, and correlations alone do not isolate net effects.

\begin{table}[htbp]
  \centering
  \caption{Pearson Correlation Matrix of Interoperability and Economic Indicators}
  \label{tab:pearson_corr}
  \scriptsize
\begin{tabular}{l c c c c c}
      \toprule
      & ASI & AAI & TVL & DAU & New Contracts \\
      \midrule
      ASI & 1.0000 & & & & \\
      \addlinespace
      AAI & \textbf{0.2276*} & 1.0000 & & & \\
          & (0.0000) & & & & \\
      \addlinespace
      TVL & \textbf{0.1625*} & \textbf{-0.1875*} & 1.0000 & & \\
              & (0.0000) & (0.0000) & & & \\
      \addlinespace
      DAU & \textbf{0.4597*} & \textbf{0.1348*} & \textbf{0.3864*} & 1.0000 & \\
              & (0.0000) & (0.0000) & (0.0000) & & \\
      \addlinespace
      New Contracts & \textbf{0.4497*} & \textbf{0.2239*} & \textbf{0.0197*} & \textbf{0.3511*} & 1.0000 \\
                    & (0.0000) & (0.0000) & (0.0072) & (0.0000) & \\
      \bottomrule
    \end{tabular}
  
  \vspace{1ex}
  \tiny
  \textit{Note:} Numbers in parentheses are p-values. * indicates significance at the 1\% level.
\end{table}

\begin{table}[htbp]
  \centering
  \small
  \setlength{\tabcolsep}{15pt}
  \begin{threeparttable}
    \caption{Heterogeneity Analysis: Impact of Interoperability by Chain Type (L1/EVM)}
    \label{tab:heterogeneity_asi_aai}
    \begin{tabular}{l c c}
      \toprule
      & (1) & (2) \\
      & TVL & Avg Gas/Tx \\
      \midrule
      ASI & -0.003 & \textbf{0.020***} \\
          & (-0.91) & (5.43) \\
      \addlinespace
      ASI\_isL1 & \textbf{-0.014***} & \textbf{-0.006***} \\
                & (-10.01) & (-5.55) \\
      \addlinespace
      ASI\_isEVM & \textbf{0.034***} & \textbf{-0.017***} \\
                 & (14.55) & (-8.78) \\
      \addlinespace
      AAI & \textbf{-9.395***} & \textbf{-4.541***} \\
          & (-13.16) & (-5.26) \\
      \addlinespace
      AAI\_isEVM & \textbf{6.568***} & \textbf{5.531***} \\
                 & (9.33) & (6.42) \\
      \addlinespace
      AAI\_isL1 & \textbf{7.163***} & \textbf{3.543***} \\
                & (12.37) & (4.61) \\
        Controls & Yes & Yes  \\
      Chain FE & Yes & Yes  \\
      Day FE   & Yes & Yes  \\ 
      \midrule
      Observations & 17,399 & 13,373 \\
      $R^2$ & 0.512 & 0.172 \\
      \bottomrule
    \end{tabular}
    
    \begin{tablenotes}[flushleft]
      \tiny
      \setlength{\itemindent}{-2.5pt}
      \item \textit{Notes:} $t$-statistics are reported in parentheses. 
      Variables with suffix \_isL1 and \_isEVM represent interaction terms with Layer-1 and EVM-compatible dummies, respectively.
      * p$<$0.1, ** p$<$0.05, *** p$<$0.01
    \end{tablenotes}
  \end{threeparttable}
\end{table}

\end{document}